\documentclass[a4paper,11pt]{article}

\usepackage[utf8x]{inputenc}
\usepackage[T1]{fontenc}
\usepackage{color}
\usepackage{amssymb}
\usepackage{amsmath}
\usepackage{graphicx}
\usepackage{mathtools}
\usepackage{ragged2e}
\usepackage{physics}

\usepackage{dcolumn}
\usepackage{bm}
\usepackage{graphicx}
\usepackage{braket}
\usepackage{verbatim} 
\usepackage[english]{babel}
\usepackage{hyperref}
\hypersetup{
citecolor=red,
colorlinks=true,
filecolor=red,
linkcolor=blue,
linktocpage=true,
urlcolor=blue
} 
\usepackage[titletoc,toc,title]{appendix}
\numberwithin{equation}{section}
\usepackage{cleveref}
\usepackage[margin=2.5cm]{geometry}
\usepackage{cite}
\usepackage{epsfig}
\usepackage{float}
\usepackage{amsfonts}
\usepackage{enumitem}
\usepackage[font={footnotesize,it}]{caption}

\numberwithin{equation}{section}

\begin{document}


\begin{center}

\centerline{\Large {\bf Mixed state entanglement in deformed field theory at finite temperature}}

\vspace{8mm}

\renewcommand\thefootnote{\mbox{$\fnsymbol{footnote}$}}
Sanjay Pant${}^{1}$\footnote{sanjaypant.phy@geu.ac.in}
and Himanshu Parihar ${}^{2, 3}$\footnote{himansp@phys.ncts.ntu.edu.tw}

\vspace{4mm}

${}^1${\small \sl Department of Allied Sciences (Physics)}\\
{\small \sl Graphic Era (Deemed to be University)}\\
{\small \sl  Dehradun, Uttarakhand 248002, India} 

\vskip 0.2cm
${}^2${\small \sl Center of Theory and Computation}\\
{\small \sl National Tsing-Hua University}\\
 {\small \sl Hsinchu 30013, Taiwan} 
 \vskip 0.2cm
${}^3${\small \sl Physics Division}\\
   {\small \sl  National Center for Theoretical Sciences}\\
     {\small \sl Taipei 10617, Taiwan} 
     \vskip 0.2cm
\end{center}

\vspace{6mm}
\numberwithin{equation}{section}
\setcounter{footnote}{0}
\renewcommand\thefootnote{\mbox{\arabic{footnote}}}

\begin{abstract}

We study mixed state entanglement measures in a higher dimensional $T\bar{T}$ deformed field theory at finite temperature. The holographic dual is described by AdS$_{d+1}$ black brane geometry with a finite cutoff. We compute the entanglement wedge cross section (EWCS), proposed to be dual to entanglement of purification (EoP) and holographic entanglement negativity (HEN) for strip like subsystems. The behavior of EWCS and HEN is studied across different regimes of temperature and deformation parameter. It is observed that the deformation and temperature exhibit similar effects on these two entanglement measures. Increasing the deformation leads to a decrease in the entanglement between the subsystems.

\end{abstract}

\newpage
\tableofcontents
\newpage

\section{Introduction}\label{sec:intro}

Quantum entanglement has recently emerged as a central theme in our understanding of various systems such as strongly coupled theories, black holes, quantum gravity, etc. In order to gain the insights of entanglement structure of such complex systems the AdS/CFT correspondence plays a crucial role \cite{Maldacena:1997re}. Quantifying entanglement or quantum correlations requires the careful analysis of measures that can accurately capture the degree of entanglement present in a given quantum state. For bipartite pure states, quantum entanglement is characterized through the entanglement entropy (EE) or the von Neumann entropy of the reduced density matrix for a particular subsystem. Although, the EE is a powerful tool for analyzing entanglement for pure states and to study the phase transition at various occasions, it is less effective in capturing the entanglement structure present in mixed states.
Several entanglement measures have been introduced within the framework of quantum information theory to characterize the entanglement present in mixed states. These measures include mutual information, entanglement of purification \cite{Terhal:2002riz}, entanglement negativity \cite{Vidal:2002zz, Plenio:2005cwa}, reflected entropy \cite{Dutta:2019gen}, and many others.
Although these quantities have been extensively studied in the context of two-dimensional conformal field theories (CFTs), directly investigating these entanglement measures for strongly coupled field theories (or CFTs) in higher dimensions remains a challenging task due to the nonperturbative nature of such theories. However, by leveraging the framework of holographic duality, we can still gain valuable insights into the entanglement structure of these higher-dimensional theories using appropriate holographic prescriptions.

A concrete example of holographic duality is the AdS/CFT correspondence which provides an intriguing connection between a $d$ dimensional CFT($CFT_d$) and quantum gravity on a $(d+1)$ asymptotically anti-de Sitter (AdS) spacetime \cite{Maldacena:1997re, Witten:1998qj}. 
In this context Ryu and Takayanagi (RT) \cite{Ryu:2006bv, Ryu:2006ef} proposed a holographic prescription to obtain the entanglement entropy for bipartite pure states in $d$ dimensional $CFT_d$'s dual to bulk asymptotic AdS$_{d+1}$ geometries. It states that the entanglement entropy of a certain region $A$ in the CFT is given by the area of a codimension two bulk static minimal hypersurface (RT surface) in the bulk AdS spacetime that is homologous to the boundary region. Later, Hubeny et al. extended this idea for time-dependent scenarios in \cite{Hubeny:2007xt}. These proposals are subsequently proved in \cite{Faulkner:2013yia,Lewkowycz:2013nqa,Dong:2016hjy}.
This has led to valuable insights into a range of phenomena from strongly correlated systems, quantum phase transitions to black hole physics \cite{Fradkin:2006mb,Nishioka:2009un,Takayanagi:2012kg,Fischler:2012ca,Fischler:2012uv,Blanco:2013joa,Chaturvedi:2016kbk,Barbosa:2024pyn}. To characterize bipartite mixed state entanglement holographically, one of the widely used quantity is entanglement wedge cross-section (EWCS) proposed in \cite{Takayanagi:2017knl, Nguyen:2017yqw} that serves as a dual of the entanglement of purification (EoP). However, a direct proof of this duality has yet to be established and its primary support comes from the fact that the properties of entanglement purification are also exhibited by the EWCS. Interestingly, the EWCS has also been proposed as a holographic dual to the odd entanglement entropy (OEE) \cite{Tamaoka:2018ned} and the reflected entropy \cite{Dutta:2019gen}. In the context of AdS$_3$/CFT$_2$, the authors in \cite{Chaturvedi:2016rcn,Chaturvedi:2017znc,Jain:2017aqk,Jain:2017uhe,Malvimat:2018txq,Malvimat:2018ood} have explored various holographic constructions to characterize the mixed state entanglement structure through entanglement negativity. These constructions involve specific algebraic sums of bulk geodesics that are homologous to appropriate combinations of subsystems and have been successful in reproducing the results obtained from dual field theory calculations \cite{Calabrese:2012ew,Calabrese:2012nk,Calabrese:2014yza,Malvimat:2017yaj} in the large central charge limit. Higher-dimensional generalizations of these constructions were proposed in \cite{Chaturvedi:2016rft,Jain:2017xsu,Jain:2018bai,KumarBasak:2020viv,Mondal:2021kzj,Afrasiar:2021hld} for a generic AdS$_{d+1}$/CFT$_d$ framework. These holographic proposals for entanglement negativity were further substantiated for subsystems with spherical entangling surfaces through the consideration of replica symmetry breaking saddles to the bulk gravitational path integral \cite{KumarBasak:2020ams, Dong:2021clv}.
Note that an alternative holographic proposal for entanglement negativity based on the bulk EWCS was also investigated in \cite{Kudler-Flam:2018qjo,Kusuki:2019zsp,KumarBasak:2020eia,KumarBasak:2021lwm}.
Interestingly the results from this proposal also match with those obtained through the aforementioned holographic proposal up to a constant for the case of AdS$_3$/CFT$_2$. See also \cite{Dong:2024gud} for recent development regarding the holographic dual of entanglement negativity.

In a separate context, the deformation of two-dimensional conformal field theories (CFT$_2$) by $T\bar{T}$ deformation \cite{Zamolodchikov:2004ce,Smirnov:2016lqw,Cavaglia:2016oda} has played a pivotal role in exploring the interplay between ultraviolet (UV) and infrared (IR) physics as well as the emergence of nonlocal effects in quantum field theories (QFRTs). The $T\bar{T}$ deformation in two-dimensional CFTs is an interesting example of a solvable irrelevant deformation. This means that despite being irrelevant under renormalization group flow, the energy spectrum and the partition function of the deformed theory can be determined exactly. The $T\bar{T}$ deformation modifies the spectrum, the entanglement structure, and the overall behavior of the original CFT$_2$ while preserving integrability and solvability. This allows for an exact determination of the energy spectrum, partition function, and S matrix \cite{McGough:2016lol}.
Generalizations have been explored including deformations involving bilinear operators constructed from conserved currents such as $J\bar{T}$ where $J$ represents a conserved holomorphic $U(1)$ current \cite{Guica:2017lia,Aharony:2018ics,Nakayama:2018ujt} and also has been extended to non relativistic systems \cite{Cardy:2018jho,Chen:2020jdi}.
Furthermore, a holographic dual for $T\bar{T}$ deformed CFTs was proposed in \cite{McGough:2016lol} where it was suggested that the deformation corresponds to introducing a cutoff in the radial direction of the dual AdS geometry. This effectively shifts the boundary of the AdS geometry toward the bulk. 
This proposal has been rigorously validated by demonstrating a remarkable agreement between holographic computations and field theory results. For example, observables such as two-point functions, energy spectrum and the partition function of the deformed CFT have been successfully reproduced within the holographic framework providing strong evidence for the consistency of this duality \cite{McGough:2016lol}. This duality offers valuable insights into the effects of the $T\bar{T}$ deformation on physical quantities within the deformed field theory
(refer to \cite{Asrat:2017tzd,Shyam:2017znq,Kraus:2018xrn,Cottrell:2018skz,Shyam:2018sro,Caputa:2019pam,Giveon:2017myj,Coleman:2020jte,Jiang:2019epa,He:2020cxp,He:2020qcs,He:2020udl,He:2022jyt,Tian:2023fgf,Dei:2024sct,Tian:2024vln} for further developments in this direction).
The entanglement entropy for bipartite pure states in $T\bar{T}$ deformed CFTs has been extensively studied in \cite{Donnelly:2018bef,Lewkowycz:2019xse,Chen:2018eqk,Banerjee:2019ewu,Jeong:2019ylz,Murdia:2019fax,Park:2018snf,Asrat:2019end,He:2019vzf,Grieninger:2019zts,Grieninger:2023knz,Banerjee:2024wtl}. It has been shown that the entanglement entropy calculated in deformed field theory matches that of holographic computations using the Ryu-Takayanagi formula \cite{Ryu:2006bv, Ryu:2006ef}.
Moreover, investigations of other bipartite mixed entanglement measures beyond entanglement entropy have been explored in \cite{Asrat:2020uib,Basu:2023bov,Basu:2023aqz,Basu:2024bal,Chang:2024voo,Basu:2024enr,Basu:2024xjq}.
A higher dimensional generalization of $T\bar{T}$ deformations have been explored in \cite{Taylor:2018xcy, Hartman:2018tkw} where the holographic dual is given by an AdS$_{d+1}$ geometry with a hard radial cutoff. The dual field theory emerges as a deformation of the higher dimensional large-N conformal field theory by an irrelevant operator which is quadratic in the stress tensor and transforms as a scalar under Lorentz transformations. This operator effectively reduces to the well-known $T\bar{T}$ deformation in the two-dimensional limit.
In relation to entanglement in higher dimensions within the context of deformed CFTs includes studies on the effect of cutoff on entanglement entropy, mutual information and EWCS in hyperscaling violating geometries at finite cutoff and zero temperature \cite{Khoeini-Moghaddam:2020ymm} and entanglement entropy in hyperscaling violating geometries at finite cutoff and finite temperature \cite{Jeong:2022jmp}. Later, the authors in \cite{Ebrahim:2023ush} investigated holographic entanglement entropy and mutual information in a deformed field theory at finite temperature dual to $(d+1)$ dimensional AdS black brane geometry with a finite cutoff.

The aforementioned developments motivate us to investigate the effect of deformation on mixed state measures like EoP and entanglement negativity in higher dimensional deformed CFTs at finite temperature. This work aims to address this issue by employing holographic techniques. 
Specifically, we investigate EoP and entanglement negativity via performing holographic computations in AdS$_{d+1}$ black brane geometry with a hard wall cutoff which is dual to deformed CFT$_d$ at finite temperature.
Following the construction of \cite{Ebrahim:2023ush}, we first analytically compute the EWCS in this deformed background. Because of the presence of nonzero deformation (finite cutoff) in addition to finite temperature, the computation of the EWCS can be obtained perturbatively in various limiting regimes of deformation parameter and temperature.
In our study we find that EWCS obeys the area law in the finite cutoff and finite temperature limits. It is also a monotonic function of temperature. We observe the decreasing nature of EWCS with respect to the deformation or cutoff parameter.
Our results reduce to the known result in the limit where the deformation vanishes, effectively recovering the finite-temperature results. We expect that this study is particularly useful as the EWCS has been shown related to the different measures of entanglement such as odd entanglement entropy, reflected entropy and entanglement negativity holographically.
Subsequently, we compute holographic entanglement negativity (HEN) for mixed state configurations of two adjacent and disjoint subsystems by employing the holographic constructions \cite{Jain:2017xsu,KumarBasak:2020viv}. Similar to the computations of EWCS, we obtain analytic expressions for HEN in various limiting regimes of temperature and nonzero deformation. We observe that deformation parameter and temperature have a similar effect on HEN.
In the limit of small deformation and low temperature, the dominant contribution comes from the AdS$_{d+1}$ vacuum and corrections to it, due to both deformation and temperature following the area law. At the high deformation regime, the volume contribution due to deformation gets completely subtracted off which is similar to the behavior observed in the high temperature limit. Furthermore, we observe that increasing the deformation parameter or temperature decreases holographic entanglement negativity and subsequently, the quantum correlations between the subsystems.
Interestingly, the results reduce to one known in literature in the limit where the deformation parameter (or cutoff) is taken to zero. This serves as a consistency check for our results.

The rest of the article is organized as follows. In \cref{review-literature} we briefly review about the construction of bulk dual to $T\bar{T}$ deformed CFTs in higher dimensions and computations of holographic entanglement entropy. In \cref{Sec-EWCS} we compute the EWCS in the present setup at different regimes of temperature and deformation parameter. Subsequently we study the effect of deformation on holographic entanglement negativity for the mixed state configuration of two adjacent and disjoint subsystems in \cref{Holographic-EN}. Finally, we conclude with a summary and discussion in \cref{summary}.

\section{Review of earlier work}\label{review-literature}

\subsection{Holographic dual of $T\bar{T}$ deformed CFT$_{d}$}	

In this subsection, we briefly review about the $T\bar{T}$ deformation in CFT$_{d}$ and its corresponding bulk dual as discussed in \cite{Ebrahim:2023ush}. In this context, consider a $d$-dimensional CFT that is deformed by an operator $X$ as
\begin{equation}\label{Deformed-CFT}
S_{\mathrm{QFT}}=S_{\mathrm{CFT}}+\lambda\int {d^dx \sqrt{\gamma}}\, X,
\end{equation}
where the operator $X$ is regarded as a generalization of the two-dimensional $T\bar{T}$ deformation to higher-dimensional CFTs \cite{Taylor:2018xcy} and $\gamma_{ij}$ is the boundary theory metric. The deformation operator $X$ in general $d$ dimensions takes the following form \cite{Taylor:2018xcy,Hartman:2018tkw}

\begin{equation}\label{TTb-bar-operator}
 X=T_{ij}~T^{ij}-\frac{1}{d-1} T^2,
\end{equation}
where $T_{ij}$ is the stress-energy tensor in the boundary theory.
The holographic bulk dual of a $T\bar{T}$ deformed CFT$_{d}$ with deformation parameter $\lambda>0$ is given by a portion of AdS$_{d+1}$ geometry with cutoff at finite radius \cite{McGough:2016lol,Taylor:2018xcy, Hartman:2018tkw}. This deformation operator $X$ is an example of a solvable irrelevant deformation with a mass dimension $2d$. The deformation parameter $\lambda$ quantifies the strength of the deformation and has a mass dimension $\Delta_\lambda =-d$. Consequently, the invariance of the effective action for such a theory gives the following flow equation

\begin{equation}\label{Flow-equation}
  \lambda~\frac{d W}{d\lambda} =\frac{1}{\Delta_{\lambda}} \int {d^dx \sqrt{\gamma}~ \langle T \rangle},
\end{equation}
where $T = \gamma^{ij} T_{ij}$ represents the trace of the field theory stress-energy tensor and $W$ denotes the on-shell effective action evaluated on the cutoff hypersurface at $z=z_c$. Comparing \cref{Flow-equation} with \cref{Deformed-CFT} gives
\begin{equation}\label{T-average}
  \langle T \rangle = - d \lambda X.
\end{equation}
The above equation implies that the deformation operator is proportional to the trace of the field theory stress-energy tensor. The bulk dual to such field theory can be described by a ($d+1$)-dimensional asymptotically AdS geometry with cutoff at $z=z_{c}$. This cutoff is related to the field theory deformation parameter $\lambda$ by the following relation:

\begin{equation}\label{lambda}
  \lambda = \frac{4 \pi G_N^{(d+1)}}{d\, R^{d-1}} z_c^d,
\end{equation}
where $R$ is the AdS radius.

According to the holographic proposal \cite{Taylor:2018xcy, Hartman:2018tkw}, the bulk geometry holographically dual to a deformed field theory at finite temperature is described by a ($d+1$)-dimensional asymptotically AdS black brane geometry with a cutoff at $z = z_c$. The metric can be expressed as
\begin{equation}\label{cutoff-metric}
ds^2=\frac{R^2}{z^2}\left( - f(z) dt^2+ d{\bold{x}}^2_{d-1}+\frac{dz^2}{f(z)} \right),
\end{equation}
where $f(z)=1-\frac{z^d}{z_h^d}$ is the blackening factor. The temperature of this black hole can be obtained by analytic continuation to a Euclidean metric,
\begin{equation}\label{BH-temp}
  T=\frac{d}{4\pi z_h}.
\end{equation}
The dual deformed field theory lives at the cutoff radius $z_c$ and the metric of the background manifold is conformal to the flat metric.

\subsection{Holographic entanglement entropy}

To define entanglement entropy in a CFT$_d$, consider a Cauchy slice in $d$-dimensional spacetime made out of two spatial subregions $A$ and $B$ such that $A\cup B$ is in a pure state $\rho_{AB}$. The entangling surface $\partial A$ represents the boundary of region $A$ and constitutes a codimension two hypersurface. The Hilbert space of this bipartite system can be described by a tensor product of the Hilbert spaces associated with the individual subsystems $A$ and $B$ as $\mathcal{H}_A \otimes\mathcal{H}_B$. The reduced density matrix corresponding to $A$ is obtained by tracing out the degrees of freedom over the subsystem $B$ as $\rho_A= \text{Tr}_B( \rho_{AB})$. Then the entanglement entropy of the subsystem $A$ is defined as the von Neumann entropy associated with this reduced density matrix by
\begin{equation}
    \mathcal{S}(A) = - \text{Tr}( \rho_A \log \rho_A).
\end{equation}
Following \cite{Ryu:2006bv, Ryu:2006ef}, the holographic entanglement entropy for bipartite states in a $d$-dimensional CFT dual to bulk asymptotically AdS$_{d+1}$ geometries is given by the area of a minimal surface homologous to the boundary subsystem as
\begin{equation}\label{RT-formula}
  S(A)=\frac{\mathcal{A}(\gamma_A)}{4G_N^{(d+1)}},
\end{equation}
where $\gamma_{\mathcal{A}}$ is a codimension two minimal surface in the bulk such that $\partial\gamma_{A}=\partial A$ and $G_N^{(d+1)}$ is the $(d+1)$-dimensional Newton's constant. 

We now briefly review the computation of holographic entanglement entropy in a deformed CFT$_d$ at finite temperature as described in \cite{Ebrahim:2023ush}. The subsystem $A$ in question is described by a $(d-1)$-dimensional spatial long rectangular strip as

\begin{equation}
  x\equiv x^1\in \left[-\frac{l}{2},\frac{l}{2}\right]~,~x^i \in  \left[-\frac{L}{2},\frac{L}{2}\right]~,i=2,...,d-1.,
\end{equation}
where $L\gg l$ in order to preserve invariance of the subsystem in $x^i$ directions. The area functional in the background geometry described in \cref{cutoff-metric} is given by 

	\begin{equation}\label{Area-surface}
	\mathcal{A}=L^{d-2} \int dz~ \left(\frac{R}{z}\right)^{d-1} \left(1-\frac{z^d}{z_h^d}\right)^{-\frac{1}{2}}\sqrt{1+x'^2 \left(1-\frac{z^d}{z_h^d}\right)}~,
	\end{equation}
where $x'=\frac{dx}{dz}$. On extremization of the above area functional, we get the following equation

	\begin{equation}\label{profile}
		\frac{dx}{dz}=\frac{\left(\frac{z}{z_*}\right)^{d-1}}{\sqrt{1-\left(\frac{z}{z_h}\right)^d}\sqrt{1-\left(\frac{z}{z_*}\right)^{2(d-1)}}}.
		\end{equation}
Here, $z=z_*$ is the turning point of the minimal surface and satisfies the condition $z'(x)=0$. Now taking the integral on both sides of \cref{profile} and further using binomial expansion leads to the following equation for subsystem size $l$ in terms of the turning point $z_*$ as

        \begin{equation}\label{length-zstar}
			\begin{aligned}
		\frac{l}{2} = z_* \sum_{k=0}^{\infty}\frac{\Gamma(k+\frac{1}{2})}{\Gamma(k+1)}&\left(\frac{z_*}{z_h}\right)^{kd}
		\bigg{[}~\frac{\Gamma\left(\frac{d(k+1)}{2(d-1)}\right)}{(1+kd)\Gamma\left(\frac{1+kd}{2(d-1)}\right)}\\
		&- \frac{1}{\sqrt{\pi} d(k+1)}\left(\frac{z_c}{z_*}\right)^{d(k+1)}~ _2F_1\left(\frac{1}{2},\frac{d(k+1)}{2(d-1)};\frac{d(3+k)-2}{2(d-1)},\left(\frac{z_c}{z_*}\right)^{2(d-1)}\right)\bigg{]}~.
		\end{aligned}
		\end{equation}
The area of the RT surface can be found by substituting \cref{profile} in \cref{Area-surface} and subsequently using a binomial expansion    

  \begin{equation}\label{area}
		\begin{aligned}
		\mathcal{A} = 2R^{d-1}&\left(\frac{L}{z_*}\right)^{d-2} \sum_{k=0}^{\infty}\frac{\Gamma(k+\frac{1}{2})}{\Gamma(k+1)} \left(\frac{z_*}{z_h}\right)^{kd}\bigg[~\frac{\Gamma\left(\frac{d(k-1)+2}{2(d-1)}\right)}{2(d-1)\Gamma\left(\frac{1+kd}{2(d-1)}\right)}\\
		&- \frac{1}{\sqrt{\pi} (d(k-1)+2)}\left(\frac{z_c}{z_*}\right)^{d(k-1)+2}~ _2F_1\left(\frac{1}{2},\frac{d(k-1)+2}{2(d-1)};\frac{d(k+1)}{2(d-1)},\left(\frac{z_c}{z_*}\right)^{2(d-1)}\right)\bigg],
		\end{aligned}
		\end{equation}
where $z_*$ is a solution to \cref{length-zstar}. To extract the leading contributions to the holographic entanglement entropy arising from the finite temperature and finite cutoff, one can perturbatively expand the aforementioned area at different limiting values of the dual field theory parameters ($l,T, \tilde{\lambda}$) as described in the subsequent subsections.

\subsubsection{Small deformation and low-temperature limit: $\tilde{\lambda}\ll l$ and $T\ll 1/l$}
\label{low small}

The small deformation and low-temperature limit is characterized by range of field theory parameters such that $Tl \ll 1$ and $\frac{\tilde{\lambda}}{l} \ll 1$ ($\tilde{\lambda}=\lambda^{1/d}$). In the bulk picture, this translates to $z_*\gg z_c$ and $z_*\ll z_h$; hence giving the following limit $z_c \ll z_* \ll z_h$. In this range, the quantities appearing in \cref{length-zstar} and \cref{area} means that the ratios $\frac{z_c}{z_*}$ and $\frac{z_*}{z_h}$ remain small, which allows the series expansion for the hypergeometric functions appearing in \cref{length-zstar} and \cref{area}. Further it is possible to write the turning point in terms of the subsystem length $l$ and cutoff $z_c$. This leads to the expression of the entanglement entropy only in terms of boundary parameters as discussed in \cite{Ebrahim:2023ush}. Finally, $z_*$ is given by

	       \begin{equation}\label{zsl}
           \begin{aligned}
	       z_* = \frac{l}{2 c_0}&\Bigg[ 1 + \frac{2}{d} (2 c_0)^{d-1} \left(\frac{z_c}{l}\right)^d-\frac{4}{d^2}\left(d-1\right)(2c_0)^{2(d-1)}
\left(\frac{z_c}{l}\right)^{2d}\\
&-\left(\frac{c_1}{(2 c_0)^{d+1}}+ \frac{c_1}{d {c_0}^2} \left(\frac{z_c}{l}\right)^d -\frac{(2 c_0)^{d-1}}{2d}\bigg(1-\frac{\left(3 d^2-7 d+6\right)c_1}{d {c_0}^2}\bigg)\left(\frac{z_c}{l}\right)^{2d}\right) \left(\frac{l}{z_h}\right)^d\\
&+\Bigg(\frac{2 {c_1}^2 \left(1+d\right) - 3 c_0 c_2}{2 (2 c_0)^{2 (d+1)}} +\frac{{c_1}^2 \left(3 d^2+7 d+6\right) - 6\left(1+d\right) c_0 c_2}{d(2 c_0)^{d + 3}}~\left(\frac{z_c}{l}\right)^{d}\\
&+\frac{1}{8 d^2 {c_0}^4}\left(d\left(d-2\right){c_0}^2 c_1
+ 2\left(5 d^2+6\right){c_1}^2-3\left(2 d^2 + d + 3\right)c_0 c_2 \right)~\left(\frac{z_c}{l}\right)^{2 d}\Bigg) \left(\frac{l}{z_h}\right)^{2 d}\Bigg],
	      \end{aligned}
	       \end{equation}
where $c_i$ are the numerical coefficient depends only on dimension $d$ is given by
\begin{equation}\label{c_i}
		 c_k \equiv \frac{\sqrt{\pi}\Gamma\left(\frac{d(k+1)}{2(d-1)}\right)}{(1+kd)\Gamma\left(\frac{1+kd}{2(d-1)}\right)},
		 \end{equation}
On using the above result in \cref{area}, the holographic entanglement entropy can be determined perturbatively as \cite{Ebrahim:2023ush}

 \begin{equation}\label{LTSD}
\begin{aligned}
  S(A)=&\frac{2R^{d-1}}{4G_N^{(d+1)}}\left(\frac{L}{z_c}\right)^{d-2}\Bigg[\frac{1}{d-2}-\frac{1}{4}\left(\frac{z_c}{z_h}\right)^{d}-\frac{3}{8\left(d+2\right)}\left(\frac{z_c}{z_h}\right)^{2d}+\mathcal{O}\left(\frac{z_c}{z_h}\right)^{3d}\Bigg]\\
  +&\frac{2R^{d-1}}{4G_N^{(d+1)}}~\left(\frac{L}{l}\right)^{d-2}\Bigg[\left(a_1+ a_5 \left(\frac{z_c}{z_h}\right)^d+a_9 \left(\frac{z_c}{z_h}\right)^{2d}+\mathcal{O}\left(\frac{z_c}{z_h}\right)^{3d}\right)\\
  +&\left(\frac{z_c}{l}\right)^d  \left(a_2+a_6 \left(\frac{z_c}{z_h}\right)^{d}+\mathcal{O}\left(\frac{z_c}{z_h}\right)^{2d}\right)+ \left(\frac{z_c}{l}\right)^{2d}\left(a_3+\mathcal{O}\left(\frac{z_c}{z_h}\right)^{d}\right)\\
  +&\left(\frac{l}{z_h}\right)^d\left(a_4+a_8 \left(\frac{z_c}{z_h}\right)^d+\mathcal{O}\left(\frac{z_c}{z_h}\right)^{2d} \right)
  +\left(\frac{l}{z_h}\right)^{2d} \left(a_7+ \mathcal{O}\left(\frac{z_c}{z_h}\right)^{d} \right)\Bigg].
    \end{aligned}
    \end{equation}
The coefficients $a_i$ detailed in the Appendix \ref{Num-coeff} are numerical factors that depend solely on the dimension $d$ similar to the $c_i$. It is observed that all these corrections follow the area law behavior. Furthermore, both temperature and the deformation parameter exert similar influences on the holographic entanglement entropy. The holographic entanglement entropy in the vanishing cutoff limit $z_c\to 0$ or $\tilde{\lambda} \to 0$ gives
\begin{equation}
  S(A)=\frac{2R^{d-1}}{4G_N^{(d+1)}}\Bigg[\left(\frac{L}{z_c}\right)^{d-2}\frac{1}{d-2}
  +\left(\frac{L}{l}\right)^{d-2}\left(a_1+a_4 \left(\frac{l}{z_h}\right)^d
  +a_7\left(\frac{l}{z_h}\right)^{2d}\right) \Bigg].
    \end{equation}
As pointed out by authors in \cite{Ebrahim:2023ush} this is the result expected for the boundary field theory as described in \cite{Fischler:2012ca}.

\subsubsection{Large deformation and low-temperature limit: $\tilde{\lambda}\gg l$ and $T\ll 1/l$}\label{lthdee}

The limit of large deformation and low temperature corresponds to a regime in the dual bulk picture where $z_c \simeq z_* \ll z_h$. In this regime, the majority of the contributions to the holographic entanglement entropy arise from the region near the cutoff $z_c$ as compared to the near horizon region. Using the convergent property of hypergeometric function $_2F_1(a,b,c;z)$ for $z=1$ when $c-a-b$ is noninteger with $\mathrm{Re}(c-a-b)>0$ one can simplify the expression of area and turning point as discussed in \cite{Ebrahim:2023ush}. Since the turning point $z_*$ is very close to cutoff $z_c$ , one can write $z_*=z_c (1+\epsilon)$ where $\epsilon = \frac{z_*-z_c}{z_c}$ is a small parameter that approaches zero as $z_*$ approaches $z_c$. On solving \cref{length-zstar} perturbatively in this small parameter $\epsilon$ leads to the following expression \cite{Ebrahim:2023ush}:

\begin{equation}\label{zs-large-def-low-temp}
\begin{aligned}
  z_*=&z_c~\Bigg[1+\frac{d-1}{8}\left(1-\left(\frac{z_c}{z_h}\right)^d~\right)~\left(\frac{l}{z_c}\right)^2\\
  +& \frac{(d-1)^2~(2d-7)}{384}~\left(1-2~ \frac{4d-7}{2d-7}~\left(\frac{z_c}{z_h}\right)^d+~ \frac{6d-7}{2d-7}~\left(\frac{z_c}{z_h}\right)^{2d}~\right)\left(\frac{l}{z_c}\right)^4\\
  +& \frac{(d-1)^3~(241-130 d+16 d^2)}{46080} \left(\frac{l}{z_c}\right)^6\\
  \times&\left(1-\frac{723-626 d+136 d^2}{241-130 d+16 d^2}~\left(\frac{z_c}{z_h}\right)^d+\frac{723-862 d+258 d^2}{241-130 d+16 d^2}~\left(\frac{z_c}{z_h}\right)^{2d}-\frac{241-366 d+138 d^2}{241-130 d+16 d^2}~\left(\frac{z_c}{z_h}\right)^{3d}\right)\Bigg].
\end{aligned}
\end{equation}
Similarly, the holographic entanglement entropy for this case may be obtained using \cref{zs-large-def-low-temp}, \cref{area} and \cref{RT-formula} as \cite{Ebrahim:2023ush}

\begin{equation}\label{LTLD}
\begin{aligned}
S_{A} = &\frac{R^{d-1}}{4G_N^{(d+1)}}\left(\frac{L}{z_c}\right)^{d-2}\bigg[\left(\frac{l}{z_c}\right)-\frac{(d-1)^2}{24}
\left(1-\left(\frac{z_c}{z_h}\right)^d\right)~\left(\frac{l}{z_c}\right)^3\\
+&\frac{(d-1)^3}{1920}(d+7)
\left(1+ 2 \frac{d-7}{d+7}\left(\frac{z_c}{z_h}\right)^d-\frac{3d-7}{d+7}\left(\frac{z_c}{z_h}\right)^{2d}\right)~\left(\frac{l}{z_c}\right)^5+\mathcal{O}~\left(\frac{l}{z_c}\right)^7\bigg].
\end{aligned}
\end{equation}
The leading term in the above above equation is the volume-dependent deformation term and the rest of the terms are corrections having the area law divergent piece at $z_c\to 0$. In the limit of large deformation and small temperature, the entanglement entropy exhibits volume law scaling analogous to the high-temperature regime where thermal entropy dominates. This implies that deformation and temperature have similar influences on the entanglement entropy. In the zero temperature limit $T\to 0$ (or $z_h \to \infty$), the above result matches the corresponding result in \cite{Khoeini-Moghaddam:2020ymm}.

\subsubsection{Small deformation and high-temperature limit: $\tilde{\lambda}\ll l$ and $1/l\ll T$}\label{sdht}
In this regime, the RT surface approaches the black hole horizon and the turning point $z_*$ is large such that $z_c \ll z_* \simeq z_h$. As described in \cite{Ebrahim:2023ush}, \cref{length-zstar} can be solved in terms of $\epsilon=\frac{z_h-z_*}{z_h}$ perturbatively after isolating the divergence and leads to the following expression \cite{Ebrahim:2023ush}

\begin{equation}\label{Epsilon-high-temp}
\begin{aligned}
  \epsilon=&\epsilon_d\exp \Bigg[-\sqrt{\frac{d(d-1)}{2}}\Biggl{\{}\frac{l}{z_h}\\
  &+2\sum_{k=0}^{\infty}\frac{1}{\sqrt{\pi}~d(k+1)}\frac{\Gamma(k+\frac{1}{2})}{\Gamma(k+1)}\left(\frac{z_c}{z_h}\right)^{d(k+1)}~ _2F_1\left(\frac{1}{2},\frac{d(k+1)}{2(d-1)};\frac{d(k+3)-2}{2(d-1)},\left(\frac{z_c}{z_h}\right)^{2(d-1)}\right)\Biggr{\}} \Bigg],
\end{aligned}
\end{equation}
where the constant $\epsilon_d$ is given by

\begin{equation}
  \epsilon_d=\frac{1}{d}\exp \left[\sqrt{\frac{d(d-1)}{2}}\Biggl{\{} 2c_0 +\sum_{k=1}^{\infty}\left(\frac{\Gamma\left(k+\frac{1}{2}\right)\Gamma\left(\frac{d(k+1)}{2(d-1)}\right)}{\Gamma\left(k+1\right)\Gamma\left(\frac{kd+1}{2(d-1)}\right)}\frac{2}{kd+1}-\sqrt{\frac{2}{d(d-1)}}\frac{1}{k}\right)\Biggr{\}}\right].
\end{equation}
Here $c_0$ is the numerical coefficient given in \cref{c_i} and the expression for $z_*$ is simply given by replacing $\epsilon$ in $z_*=z_h(1-\epsilon)$. The holographic entanglement entropy at small deformation and high temperature limit can be obtained perturbatively using $z_*$ in \cref{area} and \cref{RT-formula} as follows \cite{Ebrahim:2023ush}:

\begin{equation}\label{HTSD}
\begin{aligned}
 S(A) =\frac{2R^{d-1}}{4G_N^{(d+1)}}&\left(\frac{L}{z_h}\right)^{d-2}\Biggl[\frac{l}{2z_h}+\hat{\mathcal{S}}-\sqrt{\frac{d-1}{2d}}\epsilon
  +\frac{1}{2d}\left(\frac{z_c}{z_h}\right)^d
 +\frac{1}{8d}\left(\frac{z_c}{z_h}\right)^{2d}\Biggr]\\
+&\frac{2R^{d-1}}{4G_N^{(d+1)}}\left(\frac{L}{z_c}\right)^{d-2}\Biggl[\frac{1}{d-2}-\frac{1}{4}\left(\frac{z_c}{z_h}\right)^d-\frac{3}{8(d+2)}\left(\frac{z_c}{z_h}\right)^{2 d} \Biggr],
\end{aligned}
\end{equation}
where the constant $\hat{\mathcal{S}}$ depends only on $d$ and is given by
\begin{equation}
\hat{\mathcal{S}} = - \frac{d-1}{d-2}c_0
     + \sum_{k=1}^{\infty}\frac{\Gamma\left(k+\frac{1}{2}\right)\Gamma\left(\frac{d(k+1)}{2(d-1)}\right)}{\Gamma\left(k+1\right)\Gamma\left(\frac{kd+1}{2(d-1)}\right)} \frac{(d-1)}{(1+ k d) \left(d\left(k-1\right)+2\right)}.
\end{equation}
The first term in \cref{HTSD} is the volume of the subsystem $(V=lL^{d-2})$ while the rest of the terms are corrections and follow the area law. This behavior is similar to holographic entanglement entropy in the low temperature and large deformation limit in \cref{LTLD} where the volume term arises from the deformation. So, the leading correction term arising from finite temperature $T$ in the high temperature limit exhibits a behavior similar to the leading correction term arising from the deformation parameter $\tilde{\lambda}$.
In the zero deformation limit $z_c \to 0$, the above result reduces to the following expression

\begin{equation}
S(A) =\frac{2R^{d-1}}{4G_N^{(d+1)}} \frac{1}{d-2} \left(\frac{L}{z_c}\right)^{d-2}+\frac{2R^{d-1}}{4G_N^{(d+1)}}\left(\frac{L}{z_h}\right)^{d-2}\Biggl[\frac{l}{2z_h}+\hat{\mathcal{S}}-\sqrt{\frac{d-1}{2d}}~\epsilon\Biggr],
  \end{equation}
where the first term is the divergent term which follows the area law. It matches with the corresponding holographic entanglement entropy in the high-temperature limit \cite{Fischler:2012ca}. 

Note that there is another possible regime for field theory parameters i.e. large deformation and high temperature which corresponds to $\tilde{\lambda }T\gg 1$. The temperature of dual field theory becomes very high in the limit $\tilde{\lambda }\to 0$ and introduces complex high energy modes rendering the validity of the duality between the field theory and its gravitational dual questionable \cite{McGough:2016lol}. Having reviewed the holographic entanglement entropy, we now move to our computation of the EWCS and holographic entanglement negativity at different regimes of dual field theory parameters in the present setup.

\section{Entanglement Wedge Cross Section}\label{Sec-EWCS}
\begin{figure}
  \centering
  \includegraphics[width=1\linewidth]{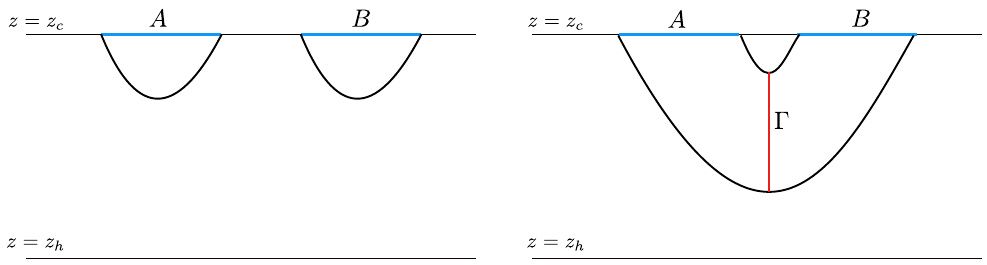}
\caption{{\bf Left:} disconnected entanglement wedge corresponding to $A$ and $B$ with vanishing EWCS. {\bf Right:} connected entanglement wedge of $A$ and $B$ with nonzero EWCS.}
\label{ewcsfg}
\end{figure}

In this section, we focus on the EWCS which serves as the holographic dual of the EoP. The EoP provides insights into the correlations within mixed states \cite{Terhal:2002riz}. To define EoP, consider a bipartite density matrix $\rho_{AB}\in \mathcal{H}_A \otimes \mathcal{H}_B$, where $\mathcal{H}_A$ and $\mathcal{H}_B$ denote the Hilbert spaces corresponding to subsystems $A$ and $B$, respectively. 
The definition of EoP requires the enlargement of Hilbert space  $\mathcal{H}_{A}\otimes \mathcal{H}_{B}\to \mathcal{H}_{A\bar{A}}\otimes \mathcal{H}_{B\bar{B}}$ by the addition of new Hilbert spaces $\mathcal{H}_{\bar{A}}$ and $\mathcal{H}_{\bar{B}}$. In this enlarged Hilbert space, the purification of $\rho_{AB}$ is defined by a pure state $\ket{\psi}\in\mathcal{H}_{A\bar{A}}\otimes \mathcal{H}_{B\bar{B}}$. Now $\rho_{AB}$ can be obtained by tracing over $\bar{A}\bar{B}$ as $\rho_{AB}=Tr_{\bar{A}\bar{B}}\ket{\psi}\bra{\psi}$. 
This method of constructing purification is not unique, as there are infinitely many possible choices for $\bar{A}$ and $\bar{B}$. The entanglement of purification, $E_p$, for $\rho_{AB}$ is then defined as
\begin{equation}
E_{p}(\rho_{AB})=\mathop{min}_{\psi,\bar{A}}S_{\rho_{A\bar{A}}}.
\end{equation}
The minimization is taken over all $\psi$ and all possible choices of partitioning $\bar{A}\bar{B}$. $S_{\rho_{A\bar{A}}}$ is the von Neumann entropy and corresponds to the reduced density matrix $\rho_{A\bar{A}}$ which can be obtained via tracing the $B\bar{B}$ part from $\ket{\psi}\bra{\psi}$. Thus, the EoP is the minimum entanglement between $A\bar{A}$ and $B\bar{B}$ in the full enlarged system. On the other hand, the EWCS which is the holographic counterpart of the EoP, is given by the minimal area of cross section of the entanglement wedge \cite{Takayanagi:2017knl,Nguyen:2017yqw}. Consider two subregions $A$ and $B$, each of length $l$ and separated by $D$ distance as shown in figure \ref{ewcsfg}. If separation $D$ is comparable or smaller than width $l$, there arises a surface denoted by $\Gamma$ in figure \ref{ewcsfg} (right). This surface connects the turning points of the RT surfaces as shown in the figure \ref{ewcsfg}. The minimal area of this $\Gamma$ surface is called the EWCS. If the separation between regions is large compared to the width of individual subregion the whole system remains in the disconnected phase and the EWCS vanishes. See \cite{BabaeiVelni:2019pkw,Jain:2020rbb,Basu:2022nds,Jain:2022hxl,Jain:2022csf, Chakrabortty:2020ptb, Karan:2023hfk,Jiang:2024akz,Mori:2024gwe,Basak:2024uwc} for recent development.

To compute the EWCS in the deformed background as given in \cref{cutoff-metric}, we specify the boundary subsystems $A$ and $B$ as follows
\begin{equation}
\begin{split}
 &  A:x^1\equiv x\in \left[ -l-\frac{D}{2},-\frac{D}{2}\right],~~ x^{j} \in \left[ -\frac{L}{2},\frac{L}{2}   \right]  \\
 &  B:x^1\equiv x\in \left[ l+\frac{D}{2},\frac{D}{2}\right],~~~~~~~ x^{j} \in \left[ -\frac{L}{2},\frac{L}{2} \right], ~~~~~~\text{where}~~j=2,\cdots, d-1.
\end{split}
\end{equation}
The induced metric on a constant time surface $\Sigma$ can be obtained from \cref{cutoff-metric} as
\begin{equation}\label{inewm}
 ds_{\Sigma_{\text{min}}}^2=\frac{R^2}{z^2}\left(d\mathbf{x}_{d-2}^2+\frac{1}{f(z)}dz^2 \right),
\end{equation}
where $\Sigma_{\text{min}}$ is the surface of minimum area positioned at $x=0$ separating $A$ and $B$. From the metric (\ref{inewm}) one can write the area functional which gives the EWCS as

\begin{equation}
\begin{split}\label{ewcsint}
E_W
&= \frac{L^{d-2}}{4G^{d+1}_N} \int_{z_1=z_{D}}^{z_{2}=z_{2l+D}} dz \left(\frac{R}{z} \right)^{d-1} \sqrt{\frac{1}{f(z)}} 
\\&= \frac{L^{d-2}R^{d-1}}{4G^{d+1}_N} \int_{z_1}^{z_2} \frac{dz}{z^{d-1}} \left( 1-\frac{z^d}{z_h^d}   \right)^{-\frac{1}{2}},
\end{split}
\end{equation}
where $z_1$ and $z_2$ are the turning points of the RT surface and corresponds to the region with width $D$ and $2l+D$, respectively. On performing the direct integration, the EWCS turns out to be proportional to hypergeometric functions of turning points
\begin{equation}
E_W
= \frac{L^{d-2}R^{d-1}}{4(d-2)G^{d+1}_N}\biggl[\frac{\, _2F_1\left(\frac{1}{2},\frac{2}{d}-1;\frac{2}{d};\frac{ {z_1^d}}{z_h^d}\right)}{{z_1}^{d-2} }-\frac{\,
   _2F_1\left(\frac{1}{2},\frac{2}{d}-1;\frac{2}{d};\frac{ {z_2}^d}{z_h^d}\right)}{{z_2}^{d-2} }\biggr].
\end{equation}

Although the integral in \cref{ewcsint} can be evaluated exactly, it is more appropriate to use the binomial expansion to make the results more tractable. The series expansion allow the study of the EWCS in different deformation and thermal limits. By doing so, the EWCS becomes
\begin{equation}\label{ewcs}
\begin{split}
 E_W&= \frac{L^{d-2}R^{d-1}}{4G^{d+1}_N} \sum_{k=0}^{\infty}  \frac{\Gamma(\frac{1}{2}+k)}{\Gamma(\frac{1}{2}) \Gamma(k+1)} \int_{z_1}^{z_2} {\left(\frac{z}{z_h} \right)^{kd} \frac{dz}{z^{d-1}}}
  \\& =\frac{L^{d-2}R^{d-1}}{4G^{d+1}_N} \sum_{k=0}^{\infty} \frac{\Gamma(\frac{1}{2}+k)}{\Gamma(\frac{1}{2}) \Gamma(k+1)}  \left(\frac{1}{kd-d+2}\right) \frac{1}{z_h^{kd}} \left(z_2^{kd-d+2}- z_1^{kd-d+2} \right).
\end{split}
\end{equation}
Our goal now is to express the EWCS in terms of the boundary parameters. This involves substituting $z_1$ and $z_2$ as a functions of the subsystem lengths into \cref{ewcs}. However, as previously discussed expressing $z_1$ and $z_2$ in terms of the boundary parameters is feasible only under specific thermal and deformation regimes. Therefore, we will focus exclusively on these limits in the following sections.

\subsection{Small deformation limit: $\tilde{\lambda}\ll D \ll l$}\label{ewcssd}
In this section we restrict ourselves to the low deformation limit which is defined by $\tilde{\lambda}\ll D \ll l$. Staying within the low deformation limit, we examine the low and intermediate thermal regimes.

\subsubsection{Small deformation and low-temperature limit: $\tilde{\lambda}\ll D \ll l$ and $T\ll \frac{1}{l},\frac{1}{D}$ }\label{ewcsll}
As discussed in the holographic entanglement entropy HEE part, the low-temperature limit is defined via $z_1<z_2\ll z_h$ or  $T\ll \frac{1}{l},\frac{1}{D}$. Because of the fact $z_1<z_2\ll z_h$ we can truncate the series in \cref{ewcs} up to the $(d+2){th}$ power of $z_1$ and $z_2$ as
\begin{equation}\label{ewl}
\begin{split}
 E_W=\frac{L^{d-2}R^{d-1}}{4G^{d+1}_N} \biggl[ \frac{\left(z_2^{-d+2}- z_1^{-d+2} \right)}{(2-d)} + \frac{\left(z_2^{2}- z_1^{2}\right)}{4 z_h^d} +\frac{3\left(z_2^{d+2}- z_1^{d+2}\right)}{8(d+2) z_h^{2d}} \biggr].
\end{split}
\end{equation}
To express the above result in terms of boundary parameters we can use \cref{zsl} for $z_1$ and $z_2$ by replacing $l$ with $D$ and $(2l+D),$ respectively. By substituting the desired values we get the EWCS in this particular limit as
\begin{equation}
   \begin{split}\label{ewll}
E_W &= E_W|_{T=0}+\frac{L^{d-2} R^{d-1}}{4G_N}\biggl\{ -\frac{l(l+D)}{4 c_0^2 z_h^d}+\frac{3(2c_0)^{-d-2}}{8(d+2)z_h^{2d}}\left((2l+D)^{d+2}-D^{d+2}\right)\\&
+\frac{D^2 c_1}{4^{d+3}c_0^3z_h^{3d}}\left(3\left(\frac{D}{c_0}\right)^{2d}+ 2^{d+2} z_h^d\left(\frac{D}{c_0}\right)^{d}+2^{2d+3} z_h^{2d}\right)-\frac{(2l+D)^2 c_1}{4^{d+3}c_0^3z_h^{3d}}\biggl[ 3\left(\frac{2l+D}{c_0}\right)^{2d} \\& 
+ 2^{d+2} z_h^d\left(\frac{2l+D}{c_0}\right)^{d} +2^{2d+3} z_h^{2d} \biggr]\biggr\}
+\frac{c_0^{d-5}z_c^d}{2^{d+7}d^2z_h^{2d}}\biggl[\mathcal{M}_1(D,z_h)+\mathcal{M}_2(2l+D,z_h)\biggr]+...,
\end{split}
\end{equation}
where $E_W|_{T=0}$ is the zero-temperature contribution to the EWCS and given by \cite{BabaeiVelni:2019pkw} 
\begin{equation}
E_W|_{T=0}= \frac{L^{d-2} R^{d-1}}{4G_N}\biggr[ \frac{(2c_0)^{d-2}}{d-2}\left(\frac{1}{D^{d-2}}-\frac{1}{(2l+D)^{d-2}}\right)\biggr].
\end{equation}
The functions $\mathcal{M}_1(D,z_h)$ and $\mathcal{M}_2(2l+D,z_h)$ are defined by

\begin{equation}
\begin{split}
   \mathcal{M}_i(x,z_h)&= \frac{1}{x^{2d-2}}\biggl[-2^d dc_0^{d+2}\left( \left( \frac{x}{z_h}\right)^d+4\right) \biggl(3\left(\frac{x}{c_0}\right)^{2d}
+2^{d+2} z_h^d \left(\left(\frac{x}{c_0}\right)^d+2^{d+1} z_h^d\right)\biggr) \\& 
+ (2c_0)^d c_1 (3d^2-7d+6) \biggl( 3\left( \frac{x}{c_0}\right)^{2d}-2^{d+2} z_h^d \left( \frac{x}{c_0}\right)^{d}+2^{d+1} z_h^d\left(\left(\frac{x}{c_0}\right)^{d}+ 2^{d+1}z_h^d\right) \biggr) \\&
+2d c_0c_1\left(\frac{x}{z_h}\right)^d \biggl[ (d+3)\left(\frac{x}{c_0}\right)^{2d}+2^{d+2}z_h^{d+2}\left( 3\left(\frac{x}{c_0}\right)^d-2^{d+1}(d-3)z_h^d\right)\biggr] \biggr].
\end{split}
\end{equation}
In \cref{ewll} the second terms inside the curly bracket are solely due to the temperature. A volume term also appears at finite temperature which is proportional to $l(l+D)$. The last term inside the square bracket that involves $\mathcal{M}_1$ is due to the deformation. As we can see by increasing the deformation at constant finite temperature the EWCS increases. On the other hand by decreasing the separation $D$ for constant $z_c$ and $l$, the EWCS shows an increasing behavior and finally diverges as $D \to 0$ as expected. 
A similar observation of the EWCS with respect to the width $D$ is documented for various other setups including AdS and deformed AdS black holes \cite{BabaeiVelni:2019pkw, Khoeini-Moghaddam:2020ymm, Chakrabortty:2020ptb}. 
 In the $z_c \to 0$ limit \cref{ewll} becomes
\begin{equation}
   \begin{split}
E_W|_{z_c\to 0} &= E_W|_{T=0}+\frac{L^{d-2} R^{d-1}}{4G_N}\biggl\{-\frac{l(l+D)}{4 c_0^2 z_h^d}+ \frac{3(2c_0)^{-d-2}}{8(d+2)z_h^{2d}}\left((2l+D)^{d+2}-D^{d+2}\right) \\&
+\frac{D^2 c_1}{4^{d+3}c_0^3z_h^{3d}}\left(3\left(\frac{D}{c_0}\right)^{2d}- 2^{d+2} z_h^d\left(\frac{D}{c_0}\right)^{d}+2^{2d+3} z_h^{2d}\right)\\&-\frac{(2l+D)^2 c_1}{4^{d+3}c_0^3z_h^{3d}}\biggl[ 3\left(\frac{(2l+D)}{c_0}\right)^{2d} - 2^{d+2} z_h^d\left(\frac{(2l+D)}{c_0}\right)^{d} +2^{2d+3} z_h^{2d} \biggr]
\biggr\}
\end{split}
\end{equation}
 The above result shows similarity with the AdS black hole results in the low-temperature regime \cite{BabaeiVelni:2019pkw}.
 
\subsubsection{Small deformation and intermediate temperature: $\tilde{\lambda}\ll D \ll l$ and $ \frac{1}{l}\ll T \ll \frac{1}{D}$ }\label{ewcsli}
Unlike the entanglement entropy, we have two distinct length scales i.e the entangling length $l$ of the subsystems and the separation between subsystems $D$. Thus, it is natural to define the small deformation and intermediate-temperature regimes by considering the high temperature and small deformation limits corresponding to $l$ and the low-temperature and small deformation limit for $D$. Therefore we can use the expression of $z_1$ from  \cref{zsl} while for the $z_2$ we can use the high temperature result as discussed in \cref{sdht}. Since $z_1 \ll z_h $ that allows us to truncate the series in $z_1$ up to the order $d+2$, while retaining the full series expression for terms involving $z_2$. 
\begin{equation}\label{ewint}
\begin{split}
 E_W= \frac{L^{d-2}R^{d-1}}{4G^{d+1}_N}\left[ \sum_{k=0}^{\infty} \frac{1}{(kd-d+2)} \frac{\Gamma(\frac{1}{2}+k)}{\Gamma(\frac{1}{2}) \Gamma(k+1)} \frac{z_2^{kd-d+2}}{z_h^{kd}}
 -\biggl( \frac{ z_1^{-d+2}}{(2-d)} + \frac{ z_1^{2}}{4 z_h^d} +\frac{ 3z_1^{d+2}}{8(d+2) z_h^{2d}} \biggr)\right]
\end{split}
\end{equation}
The first term in the above equation is not convergent for large value of $k$ due to $\frac{z_2}{z_h} \to 1$. In the large $k$ limit, one can approximate this term by
\begin{equation}
    \sum_{k=0}^{\infty}\frac{1}{k}\left(\frac{z_2}{z_h}\right)^{kd}=-\ln{\left[1-\left(\frac{z_2}{z_h}\right)^{d}\right] }
\end{equation}
 Using the analysis done in \cref{Epsilon-high-temp} for the high temperature case, \cref{ewint} is
\begin{equation}
\begin{split}
 E_W= \frac{L^{d-2}R^{d-1}}{4G^{d+1}_N}\left[ \sum_{k=0}^{\infty} \frac{1}{(kd-d+2)} \frac{\Gamma(\frac{1}{2}+k)}{\Gamma(\frac{1}{2}) \Gamma(k+1)} \frac{1- (kd-d+2)\epsilon}{z_h^{d-2}}
 -\biggl( \frac{ z_1^{-d+2}}{(2-d)} + \frac{ z_1^{2}}{4 z_h^d} +\frac{ 3z_1^{d+2}}{8(d+2) z_h^{2d}} \biggr)\right]
\end{split}
\end{equation}
In the large $k$ approximation the above equation gives
\begin{equation}
\begin{split}
 E_W &= \frac{L^{d-2}R^{d-1}}{4G^{d+1}_N}\left[ -\ln{\left[1-\left(\frac{z_2}{z_h}\right)^{d}\right]}
 -\biggl( \frac{ z_1^{-d+2}}{(2-d)} + \frac{ z_1^{2}}{4 z_h^d} +\frac{3 z_1^{d+2}}{8(d+2) z_h^{2d}} \biggr)\right]
\end{split}
\end{equation}
Substituting $z_1$ in terms of $\epsilon$ from \cref{Epsilon-high-temp} we obtain the final expression for EWCS as
\begin{equation}\label{ewfint}
\begin{split}
E_W &= E_W|_{T=0}+ \frac{L^{d-2}R^{d-1}}{4G^{d+1}_N}\biggl\{-\sqrt{\frac{d(d-1)}{2}}\biggl[ \frac{2l+D}{z_h}+2\sum_{k=0}^{\infty} \frac{1}{\sqrt{\pi}d(k+1)} \frac{\Gamma(k+\frac{1}{2})}{ \Gamma(k+1)} \left(\frac{z_c}{z_h}\right)^{d(k+1)} \\&
_2F_1\left(\frac{1}{2},\frac{d(k+1)}{2(d-1)};\frac{d(k+3)-2}{2(d-2)};\left(\frac{ z_c}{z_h}\right)^{2(d-1)}\right)\biggr]- \biggl[
\frac{3}{2^{d+5}(d+2)z_h^{2 d}}\left(\frac{D }{c_0}\right)^{d+2} \\&
+\frac{D ^2 }{16 z_h^{d} c_0^2}-\left(\frac{D}{z_h}\right)^d \left(\frac{3c_1D^{d+2}}{2^{2d+6}z_h^{2d}c_0^{2d+3}}+\frac{D^2 c_1}{2^{d+4}z_h^{d} c_0^{d+3}} +\frac{c_1}{8c^3D^{d-2}}\right)+\left(\frac{z_c}{D}\right)^d 
\biggl(\frac{3 D^{d+2}}{32d z_h^{2d}c_0^3}\\&+\frac{2^{2d-2}}{dc_0^3D^{d-2}}
+\frac{2^{d-3} D^2 c_0^{d-3}}{d z_h^{d}}
+\left(\frac{D}{z_h}\right)^d \biggl(-\frac{c_0^{2d-5}}{2^{d+7}d^2 D^{d-2} z_h^{2 d}} \biggl(c_1 2^d (d(3d-7)+6) c_0^d\biggl(3 \left(\frac{D}{c_0}\right)^{2d}\\&+2^{d+2} z_h^d \left(\left(\frac{D}{c_0}\right)^d
+2^{d+1} z_h^d\right)\biggr)-2^d d c_0^{d+2} \left(3 \left(\frac{D}{c_0}\right)^{2d}+2^{d+2} z_h^d \left(\left(\frac{D}{c_0}\right)^d+2^{d+1} z_h^d\right)\right)\\&
+2 c_1 c_0 d \biggl(3(d+3) \left(\frac{D}{c_0}\right)^{2 d}+2^{d+2} z_h^d \left(3 \left(\frac{D}{c_0}\right)^d-2^{d+1}(d-3)z_h^d\right)\biggr)\biggr)\biggr)\biggr)\biggr]\biggr\},
\end{split}
\end{equation}
where the zero-temperature part is
\begin{equation}
E_W|_{T=0} = \frac{L^{d-2}R^{d-1}}{4G^{d+1}_N}\biggl[-\ln{(\epsilon_dd)}-\frac{2^{d-2}}{(2-d)} \left(\frac{c_0 }{D}\right)^{d-2}\biggr].
\end{equation}
The \cref{ewfint} shows a UV divergence near the boundary of spacetime as $D\to 0$.
The contribution due to the cutoff arises from the terms involving $z_c$. The second term $E_W|_{T=0}$ follow the area law. The correction term in \cref{ewfint} that involves the summation follows the area law. The above result indicates that the corrections to the EWCS due to the deformation parameter manifest in two distinct ways. First, through terms proportional to $\frac{z_c}{z_h}$ and, second through terms that depend solely on variations in the width $D$. Similar observations were reported previously for mutual information which provides lower bound on the EWCS \cite{Ebrahim:2023ush}. It is clear that EWCS follows the area law at the intermediate temperature that makes it more relevant measure for the mixed state entanglement unlike the HEE which scales only with volume in this limit. As the cutoff increases, the value of the EWCS decreases which indicates the decrease in total correlations. The area law behavior is generally associated with the quantum correlations, therefore the decrease in the EWCS with deformation in \cref{ewfint} suggests a decline in quantum entanglement between subregions $A$ and $B$.

Further, in the $z_c\to 0$ limit \cref{ewfint} reduces to the EWCS for the AdS black brane geometry up to some terms as obtained in \cite{BabaeiVelni:2019pkw}.
\begin{equation}
\begin{split}
E_w|_{z_c\to 0} =E_W|_{T=0}+ \frac{L^{d-2}R^{d-1}}{4G^{d+1}_N}&\biggl[-\sqrt{\frac{d(d-1)}{2}}\biggl(\frac{2l+D}{z_h}\biggr)- \biggl[
\frac{D ^2 }{16 z_h^{d}c_0^2}+\frac{3}{2^{d+5}(d+2)z_h^{2 d}}\left(\frac{D }{c_0}\right)^{d+2} \\&
-\left(\frac{D}{z_h}\right)^d \left(\frac{3c_1D^{d+2}}{2^{2d+6}z_h^{2d}c_0^{2d+3}}+\frac{D^2 c_1}{2^{d+4}z_h^{d} c_0^{d+3}} +\frac{c_1}{8c^3D^{d-2}}\right)\biggr]\biggr].
\end{split}
\end{equation}
In the small deformation regime we observe that the EWCS follows the area law at low and intermediate thermal limits and consists of the correction terms due to the deformation. The deformation or cutoff parameter appears in two different combinations. In the upcoming subsection, we deal with the intermediate range of deformation or cutoff that is very close to the turning point of the RT surface corresponding to the subsystem of width $D$.

\subsection{ Intermediate deformation limit:  $D \ll \tilde{\lambda} \ll l $ }\label{ewcsi}
The intermediate range of deformation is defined by taking the cutoff $z_c$ close to the turning point $z_1$. In this range $z_c \to z_1$ prevents us from taking $z_c$ to zero limit. However, we can still explore both the low and high temperature limits. Since $z_c$ is comparable to $z_1$ therefore both turning points behave differently with respect to the cutoff. This gives rise to two different thermal limits as discussed below.

\subsubsection{Intermediate deformation and low-temperature limit: $D \ll \tilde{\lambda} \ll l $ and $T\ll \frac{1}{l},\frac{1}{D}$}\label{ewcsil}
This limit corresponds to the low temperature for both the turning points along with the condition that $z_c \to z_1$. The EWCS in this limit can be computed by using the low temperature and high deformation expression of $z_1$ discussed in \cref{lthdee} and the low temperature and small deformation expression of $z_2$ given in \cref{zsl}. Because of the complexity of computation we only consider the terms up to the $(\frac{D}{z_c})^4$ of \cref{zs-large-def-low-temp}. Now using \cref{ewcs} we get the following form of the EWCS:

\begin{equation}\label{idltmp}
    \begin{split}
        E_W&=\frac{L^{d-2}R^{d-1}}{4G^{d+1}_N}\biggl\{\frac{1}{2-d}\biggl[\frac{1}{8 dc_0^4 (D+2l)^{d-2}}  \biggl(-(2c_0)^dc_1 \left(d^2-5d+6\right) \left(\frac{z_c}{z_h}\right)^d
   +dc_0 c_1 (d-2) \left(\frac{D +2l}{z_h}\right)^d \\&-(2c_0)^{2d+1} (d-2) \left(\frac{z_c}{D +2
   l}\right)^d+2^{d+1} d c_0^{d+2}\biggr)
   -\frac{1}{z_c^{d-2}} \biggl\{1+\frac{(2-d) (d-1)}{8} \left(1-\left(\frac{z_c}{z_h}\right)^d\right) \left(\frac{D
   }{z_c}\right)^2 \\&
   +\frac{(2-d) (d-1)^2 (2 d-7)}{384} 
   \mathcal{K}(z_c, z_h, d) \left(\frac{D}{z_c}\right)^4\biggr\}\biggr]  
    \\& +\frac{3}{8 (d+2)z_h^{2d}} \biggl[\frac{(D +2 l)^{d+2}}{2^{d+3}dc_0^{d+4}} \biggl(-\frac{c_1 d(d+2)}{2^d c_0^{d-1}} \left(\frac{D +2l}{z_h}\right)^d 
   -(d+2) \left(\frac{z_c}{D +2 l}\right)^d  \\& \left(c_1 (d+3) \left(\frac{D +2 l}{z_h}\right)^d-(2c_0)^{d+1}\right)+2 c_0^2 d\biggr) 
   -z_c^{2+d}
   \biggl\{1+\frac{(d+2) (d-1)}{8} \left(1-\left(\frac{z_c}{z_h}\right)^d\right) \left(\frac{D }{z_c}\right)^2 \\&
   +\frac{(d-1)^2 (d+2) (2 d-7)}{384}\mathcal{K}(z_c, z_h, d) \left(\frac{D }{z_c}\right)^4\biggr\}\biggr]
   +\frac{1}{16dc_0^4 z_h^{d}} \biggl((D +2l)^2 \biggl(c_0^2 d-\frac{c_1d}{2^dc_0^{d-1}}\left(\frac{D +2l}{z_h}\right)^d \\&
   +\left(\frac{z_c}{D +2l}\right)^d \left((2c_0)^{d+1}
   -3 c_1 \left(\frac{D +2 l}{z_h}\right)^d\right)\biggr)
   -4 c_0^4 d z_c^2 \biggl(1+\frac{1}{4} (d-1)
   \left(1-\left(\frac{z_c}{z_h}\right)^d\right) \left(\frac{D }{z_c}\right)^2 \\&
   +\frac{(d-1)^2 (2 d-7)}{192} \mathcal{K}(z_c, z_h, d) \left(\frac{D}{z_c}\right)^4\biggr)\biggr)\biggr\}.
    \end{split}
\end{equation}
where the function $\mathcal{K}(z_c, z_h, d)$ is
\begin{equation}
    \mathcal{K}(z_c, z_h, d)= \biggl(1-\frac{2 (4d-7)
   }{2d-7}\left(\frac{z_c}{z_h}\right)^d
   +\frac{(6 d-7)}{2 d-7}\left(\frac{z_c}{z_h}\right)^{2d}\biggr).
\end{equation}
Note that similar to the small deformation case corrections to the EWCS arise independently from both the temperature and the deformation parameter. We observe that, as deformation increases, the EWCS decreases. Since we are working in the finite nonzero deformation range that restrict us from take the vanishing $D$ limit. Instead we can examine the vanishing temperature limit $T\to 0$ of the equation
\begin{equation}
    \begin{split}
        E_W|_{T\to 0}&=\frac{L^{d-2}R^{d-1}}{4G^{d+1}_N}\frac{1}{2-d}\biggl[\frac{1}{8 dc_0^4 (D+2l)^{d-2}}  \biggl(-(2c_0)^{2d+1} (d-2)  \left(\frac{z_c}{D +2l}\right)^d+2^{d+1} c_0^{d+2}d\biggr)\\&
   -z_c^{2-d} \biggl(1+\frac{(2-d) (d-1)}{8} \left(\frac{D
   }{z_c}\right)^2 +\frac{(2-d) (d-1)^2 (2 d-7) }{384}\left(\frac{D}{z_c}\right)^4\biggr)\biggr].
    \end{split}
\end{equation}
In the $T\to 0$ limit we can see that as the deformation or cutoff increases, the EWCS decreases.

\subsubsection{Intermediate deformation and intermediate temperature: $D \ll \tilde{\lambda} \ll l $ and $ \frac{1}{l}\ll T \ll \frac{1}{D}$ }\label{ewcsii}
In the intermediate deformation and intermediate temperature limit, we directly substitute $z_1=z_c$ while using the low-temperature limit expression for $z_2$ given in \cref{zsl}. We can use \cref{ewl} since both the turning points are still very small as compared to the horizon. Finally the EWCS for this case is
\begin{equation}
\begin{split}
E_W&= \frac{L^{d-2}R^{d-1}}{4G^{d+1}_N}\biggl\{\biggl(\frac{3 \left(\left(\frac{D +2 l}{2c_0}\right)^{d+2}-z_c^{d+2}\right)}{8z_h^{2 d}(d+2)}+\frac{1}{4z_h^{d}} \left(\frac{(D +2l)^2}{4 c_0^2}-z_c^2\right)-\frac{2^{2-d} c_0^{d-2} (D
+2 l)^{2-d}}{d-2}+\frac{z_c^{2-d}}{d-2}\\&
\frac{c_0^{2 d-3} z_c^d }{32dz_h^{2d}(D +2l)^{2d-2}} \left(2^{d+2} z_h^d \left(\left(\frac{D +2l}{c_0}\right)^d+2^{d+1} z_h^d\right)+3 \left(\frac{D +2l}{c_0}\right)^{2d}\right)+..\biggr)\\&
+\biggl[-\frac{c_1}{4^{d+3}(D +2 l)^{d-2}z_h^{2 d}c_0^3}\left(2^{d+2} z_h^d \left(\left(\frac{D +2l}{c_0}\right)^d+2^{d+1} z_h^d\right)+3 \left(\frac{D +2l}{c_0}\right)^{2 d}\right)\\&
   +\frac{c_0^{d-5}}{2^{d+7}d^2 (D +2l)^{d-2} z_h^{2d}}\biggl( 3\left(\frac{D+2l}{c_0}\right)^{2d}\biggl(-2d(d+3)c_0c_1+2^d c_0^d(dc_0^2+(7d-3d^2-6)c_1))\biggr)\\&+2^{d+2}\left(\frac{D+2l}{c_0}\right)^d\left(-6dc_0c_1+2^dc_0^d(-3d^2+7d-6)c_1\right)z_h^d +2^{2d+3}\biggl(2(d-3)dc_0c_1\\&
   +2^dc_0^2\left(dc_0^2(-6+(7-3d)d)c_1\right)\biggr)z_h^{2d}\biggr)\left(\frac{z_c}{D+2l}\right)^d+..\biggr]\biggl(\frac{D+2l}{z_h}\biggr)^{d}+....\biggr\}.
\end{split}
\end{equation}
  Similar to the previous section it is observed that the EWCS decreases with deformation. It remains finite at zero separation $D\to 0$ between the subsystems. However, $D\to 0$ is not a valid limit for this particular regime.
By simplifying the second term of the above results, we can observe that it is proportional to the area, thereby following the area law. A careful analysis of area law behavior suggests a decrease in entanglement due to the deformation that is similar to the effect of increasing temperature on entanglement.

So far we have seen that in the small and intermediate deformation range the EWCS decreases with deformation and obeys area law. These results indicate that the presence of finite cutoff or deformation reduces the correlations. In the next section we discuss the high deformation limit that allows us to explore the low-temperature regime only. 

\subsection{ High deformation limit:  $D \ll l \ll \tilde{\lambda}$ }\label{ewcsh}
In the high deformation regime we only have the low-temperature case as mentioned earlier. We can write the required relation of turning points from \cref{zs-large-def-low-temp}.
The final form of EWCS after substituting \cref{zs-large-def-low-temp} in \cref{ewcs} is given by
\begin{equation}\label{hdf}
\begin{split}
E_W&=\frac{L^{d-2}R^{d-1}(d-1)}{4G^{d+1}_N}\biggl\{\frac{4l(l+D)}{z_c^{d}} \biggl[\frac{1}{8} \left(1-\left(\frac{z_c}{z_h}\right)^d\right)+\frac{(d-1) (2 d-7)}{384}\left(\left(\frac{D }{z_c}\right)^2+\left(\frac{2 l+D }{z_c}\right)^2\right) \\&\left(1-\frac{2 (4 d-7) }{2 d-7}\left(\frac{z_c}{z_h}\right)^d+\frac{(6 d-7)}{2 d-7}\left(\frac{z_c}{z_h}\right)^{2 d}\right) \biggr]
+\frac{3l(l+D) z_c^{d}}{2z_h^{2d}}\bigg[\frac{1}{8} \left(1-\left(\frac{z_c}{z_h}\right)^d\right) \\&
+\frac{(d-1) (2 d-7)}{384} \left(1-\frac{2 (4 d-7)}{2 d-7}\left(\frac{z_c}{z_h}\right)^d+\frac{(6 d-7)}{2 d-7}\left(\frac{z_c}{z_h}\right)^{2 d}\right)
   \left(\left(\frac{2l+D }{z_c}\right)^2+\left(\frac{D }{z_c}\right)^2\right)\biggr]\\&
   +\frac{l(l+D)}{z_h^{d}}\biggl[\frac{1}{4} \left(1-\left(\frac{z_c}{z_h}\right)^d\right) +\frac{(d-1) (2 d-7)}{192}
    \biggl(1-\frac{2 (4d-7)}{2 d-7}\left(\frac{z_c}{z_h}\right)^d+\frac{(6d-7)}{2 d-7}\left(\frac{z_c}{z_h}\right)^{2 d}\biggr) \\&\biggl(\left(\frac{2
   l+D}{z_c}\right)^2+\left(\frac{D}{z_c}\right)^2\biggr)\biggr]\biggr\}.
   \end{split}
\end{equation}
 The ratio $z_c/z_h \ll 1,$ therefore we can neglect the higher order contributions. In the result above, we observe that, unlike the previous cases, the cutoff parameter appears in the denominators. It is important to note that we are working in the high deformation limit, which prevents us from taking the limit $z_c \to 0$. As a result, the EWCS does not diverge with vanishing cutoff. Also the EWCS decreases with increasing deformation at constant temperature.
In the zero-temperature limit, we have
\begin{equation}\label{hdf0t}
\begin{split}
E_W|_{T\to 0}&=\frac{L^{d-2}R^{d-1}}{4G^{d+1}_N}\frac{4l(l+D)(d-1)}{z_c^{d}} \biggl[\frac{1}{8}+\frac{ (d-1) (2 d-7)}{384} \left(\left(\frac{D }{z_c}\right)^2+\left(\frac{2 l+D }{z_c}\right)^2\right)\biggr].
   \end{split}
\end{equation}
It is clear that the EWCS is a decreasing function of the cutoff at zero temperature similar to the observation in \cite{Khoeini-Moghaddam:2020ymm}.

Throughout in our analysis we saw how the deformation and the temperature affects the mixed state entanglement by studying its effect on the EWCS. We also observed that the EWCS follows area law unlike the entanglement entropy for high thermal and deformation limits \cite{Ebrahim:2023ush}.
It is important to note that, although the EWCS measures total correlation (classical and quantum), we observe that corrections to the EWCS in the present setup scales as area law in different deformation and thermal limits. Also, increasing deformation decreases the EWCS and thus one can conclude that deformation reduces the entanglement between the subsystems. The deformation and temperature have a similar effect as both tend to reduce the EWCS. This behavior appears to suggest that increasing these parameters introduces additional degrees of freedom (like thermal correlations for the case of temperature) within the system that leads to a decrease in entanglement.
In the following, we focus on another mixed state entanglement measure called entanglement negativity which only quantifies the quantum correlation in a given mixed state.

\section{Holographic Entanglement Negativity}\label{Holographic-EN}

In this section, we utilize the holographic proposals discussed in \cite{Jain:2017xsu,KumarBasak:2020viv} to obtain the holographic entanglement negativity analytically for the bipartite mixed state configurations of two adjacent and disjoint subsystems in a $T\bar{T}$ deformed field theory at finite temperature. These proposals involve an algebraic sum of the areas of the bulk codimension two extremal surfaces in the bulk homologous to the relevant subsystems.
Note that there exists another holographic proposal for entanglement negativity based on the EWCS \cite{Kudler-Flam:2018qjo,Kusuki:2019zsp} as mentioned in the Introduction \cref{sec:intro}, and both proposal reproduces the universal logarithm term in two-dimensional CFTs. We use the former proposal in our computations.

We first briefly review the definition of entanglement negativity in the context of quantum information theory. Consider a tripartite system in a pure state that consists of the subsystems $A_1$, $A_2$ and $B$ such that $A\equiv A_1\cup A_2$ and $B= A^c$ represents the rest of the subsystem. 
The relevant Hilbert space is given by $\mathcal{H}_A=\mathcal{H}_{A_1}\otimes \mathcal{H}_{A_2}$, where $\mathcal{H}_{A_1}$ and $\mathcal{H}_{A_2}$ denote the Hilbert
space for the subsystems $A_{1,2}$ respectively. The partial transpose $\rho_A^{T_{A_2}}$ of the reduced density matrix $\rho_A$ with respect to the subsystem $A_2$ is defined by

\begin{equation}
\mel{e^{(A_1)}_ie^{(A_2)}_j}{\rho_A^{T_{A_2}}}{e^{(A_1)}_ke^{(A_2)}_l}=\mel{e^{(A_1)}_ie^{(A_2)}_l}{\rho_A}{e^{(A_1)}_ke^{(A_2)}_j},
\end{equation}
where $\mathcal{H}_{A_1}$ and $\mathcal{H}_{A_2}$ are the orthonormal bases for the Hilbert spaces $\mathcal{H}_{A_1}$ and $\mathcal{H}_{A_2}$ respectively. The entanglement negativity $\mathcal{E}$ for the bipartite mixed state configuration of subsystems $A_1$ and $A_2$ may then be defined as the logarithm of the trace norm of the partially transposed reduced density matrix as
\begin{equation}
\mathcal{E} = \ln\mathrm{Tr}\left|\rho_A^{T_{A_2}}\right|,
\end{equation}
where the trace norm $\mathrm{Tr}|\rho_A^{T_{A_2}}|$ denote the sum of absolute eigenvalues of $\rho_A^{T_{A_2}}$. Entanglement negativity is additive, monotonic and obeys monogamy; however, it is not convex and characterizes an upper bound on the distillable entanglement of the mixed state.

\subsection{Adjacent subsystems}

We now proceed with the mixed state configuration of two adjacent subsystems $A_1\cup A_2$  in a dual CFT$_d$ at finite temperature with deformation. These subsystems are described by the rectangular strip geometries as
\begin{equation}
x\equiv x^1=\biggl[-\frac{l_j}{2},\frac{l_j}{2}\biggr],\quad x^i=\biggl[-\frac{L}{2},\frac{L}{2}\biggr];\quad i=2,3,\dots,(d-1),\quad j=1,2,
\end{equation}
where $L\gg l_1,l_2$ is taken to preserve translational invariance of the subsystems in $x^i$ directions. Following \cite{Jain:2017xsu}, the holographic entanglement negativity between two adjacent subsystems is given by
\begin{equation}\label{HEN-adjacent-formula}
\begin{aligned}
    \mathcal{E}&=\frac{3}{16G_N^{(d+1)}}(\mathcal{A}_{1}+\mathcal{A}_{2}-\mathcal{A}_{12}),\\
    &=\frac{3}{4}(S(A_1)+S(A_2)-S(A_1\cup A_2)),
        \end{aligned}
\end{equation}
where $\mathcal{A}_{i}$ and  $S(A_i)$ denote the area of the RT surface and holographic entanglement entropy corresponding to the subsystem $A_i$, respectively. The second line in the above equation can be written using the RT formula \cite{Ryu:2006bv}.  For the configuration of adjacent subsystems, we have four parameters in the dual field theory i.e $l_1,l_2, \tilde{\lambda}$ and $T$, so we have corresponding three different regimes for these parameters. These are: \\(a) Small deformation and low-temperature regime where $\tilde{\lambda}  \ll l_{1,2}$ and $T l_{1,2} \ll 1$. \\(b) small deformation and high-temperature regime, where $ \tilde{\lambda}\ll l_{1,2}$ and $ T l_{1,2}\gg 1$. \\(c) large deformation and low temperature regime where $l_{1,2} \ll \tilde{\lambda}$ and $T l_{1,2}\ll 1$. \\Large deformation and high-temperature limit is not possible due to the presence of complex high energy modes as described earlier. Note that we are taking $l_1$ and $l_2$ to be of the same order. Also note that the horizon imposes a fundamental constraint on the cutoff scale in all these regimes leading to $\tilde{\lambda} T \ll 1$ and is equivalent to $z_c/z_h \ll 1$ in the bulk picture. We now employ the above holographic proposal to compute the entanglement negativity for two adjacent subsystems in the present setup in the subsequent subsections.

\subsubsection{Small deformation and low-temperature regime: $\tilde{\lambda}  \ll l_{1,2}$ and $T \ll 1/l_{1,2}$}
In the regime of small deformation and low temperature in the deformed theory at finite temperature, we can use \cref{LTSD} for $S(A_1)$, $S(A_2)$ and also $S(A_1 \cup A_2)$ in \cref{HEN-adjacent-formula} to obtain the holographic entanglement negativity for the mixed state configuration of adjacent subsystems as follows

\begin{equation}\label{EN-ajd-small-def-low-temp}
\begin{aligned}
  \mathcal{E}=&\frac{3R^{d-1}}{8 G_N^{(d+1)}}\left(\frac{L}{z_c}\right)^{d-2}\Bigg[\frac{1}{d-2}+a_1 \mathcal{F}_{d-2}^{\mathrm{adj}} -\frac{1}{4}\left(\frac{z_c}{z_h}\right)^{d}-\frac{3}{8\left(d+2\right)}\left(\frac{z_c}{z_h}\right)^{2d}\\
  &\hspace{1.5cm}+\left(a_5 \left(\frac{z_c}{z_h}\right)^d+a_9 \left(\frac{z_c}{z_h}\right)^{2d}\right)\mathcal{F}_{d-2}^{\mathrm{adj}} 
  +\left(a_2+a_6 \left(\frac{z_c}{z_h}\right)^{d}\right)\mathcal{F}_{2d-2}^{\mathrm{adj}}+ a_3 \mathcal{F}_{3d-2}^{\mathrm{adj}}\Bigg]\\
&+\frac{3R^{d-1}}{8G_N^{(d+1)}}~\left(\frac{L}{z_h}\right)^{d-2}\Bigg[\left(a_4+a_8 \left(\frac{z_c}{z_h}\right)^d \right)\mathcal{G}_2^{\mathrm{adj}}
  +a_7 \mathcal{G}_{d+2}^{\mathrm{adj}}\Bigg]+...,
    \end{aligned}
    \end{equation}
where $\mathcal{F}_n^{\mathrm{adj}} = z_c^n \left(\frac{1}{l_1^n}+ \frac{1}{l_2^n} - \frac{1}{(l_1+l_2)^n}\right)$, $\mathcal{G}_n^{\mathrm{adj}} = z_h^{-n} \left(l_1^n+l_2^n - (l_1+l_2)^n\right),$ and the ellipsis denotes the higher order terms. The first term in the above result is UV divergent in the limit $z_c \to 0$. On taking the limit $z_c\to 0$ and $T\to 0$ in \cref{EN-ajd-small-def-low-temp} gives

\begin{equation}
\begin{aligned}
  \mathcal{E}_{z_c\to 0,T\to 0}=\frac{3R^{d-1}}{8 G_N^{(d+1)}}\Bigg[\frac{1}{d-2}\left(\frac{L}{z_c}\right)^{d-2}+a_1 \left(\frac{1}{l_1^{d-2}}+ \frac{1}{l_2^{d-2}} - \frac{1}{(l_1+l_2)^{d-2}}\right)\Bigg].
    \end{aligned}
    \end{equation}
where $z_c$ is the UV cutoff in the $z_c\to 0$ limit. Interestingly, the above result matches with the holographic entanglement negativity for two adjacent subsystems in a CFT$_d$ at zero temperature dual to AdS$_{d+1}$ vacuum \cite{Jain:2017xsu}. In \cref{EN-ajd-small-def-low-temp}, additional terms beyond the vacuum contribution represent corrections due to finite temperature and finite cutoff (or deformation) to the holographic entanglement negativity. The overall contribution of these corrections are negative and hence they reduce the entanglement negativity between the subsystems. Note that this behavior is similar to that of mutual information in small deformation and low temperature limit \cite{Ebrahim:2023ush} which is expected as holographic entanglement negativity is proportional to the mutual information\footnote{It is important to note that the entanglement negativity and mutual information are different measures in quantum information theory. Nevertheless, for the bipartite configuration considered, their universal part coincides and it becomes dominant in the holographic (large central charge) limit.} for the case of adjacent subsystems.

If we specifically consider the zero deformation limit where $z_c\to 0$ in \cref{EN-ajd-small-def-low-temp}, we obtain the following expression that incorporates only finite temperature corrections

\begin{equation}
\begin{aligned}
  \mathcal{E}_{z_c\to 0}=&\frac{3R^{d-1}}{8 G_N^{(d+1)}}\Bigg[\frac{1}{d-2}\left(\frac{L}{z_c}\right)^{d-2}+a_1\left(\frac{1}{l_1^{d-2}}+ \frac{1}{l_2^{d-2}} - \frac{1}{(l_1+l_2)^{d-2}}\right)\Bigg]\\
&+\frac{3R^{d-1}}{8G_N^{(d+1)}}\left(\frac{L}{z_h}\right)^{d-2}\Bigg[a_4 \mathcal{G}_2^{\mathrm{adj}}
  +a_7 \mathcal{G}_{d+2}^{\mathrm{adj}}\Bigg].
    \end{aligned}
    \end{equation}
The above result matches with the holographic entanglement negativity for adjacent subsystems in the low-temperature limit obtained in \cite{Jain:2017xsu} up to order $1/z_h^d$. This matching serves as a consistency check for our computation.

On the other hand, to specifically examine the impact of deformation on entanglement negativity at zero temperature, we can take the limit as the temperature $T\to 0$ in \cref{EN-ajd-small-def-low-temp} as
\begin{equation}
\begin{aligned}
  \mathcal{E}_{T\to 0}=&\frac{3R^{d-1}}{8 G_N^{(d+1)}}\Bigg[\frac{1}{d-2}\left(\frac{L}{z_c}\right)^{d-2}+a_1\left(\frac{1}{l_1^{d-2}}+ \frac{1}{l_2^{d-2}} - \frac{1}{(l_1+l_2)^{d-2}}\right)\Bigg]\\
  +&\frac{3R^{d-1}}{8 G_N^{(d+1)}}L^{d-2}\Bigg[a_2 z_c^d \left(\frac{1}{l_1^{2d-2}}+ \frac{1}{l_2^{2d-2}} - \frac{1}{(l_1+l_2)^{2d-2}}\right)+a_3 z_c^{2d}\left(\frac{1}{l_1^{3d-2}}+ \frac{1}{l_2^{3d-2}} - \frac{1}{(l_1+l_2)^{3d-2}}\right)\Bigg].
    \end{aligned}
    \end{equation}
Here, the first line is the usual vacuum result and terms in second line represent the correction due to finite cutoff. These correction terms are proportional to the area of the entangling surface ($A=L^{d-2}$) between the subsystems similar to the finite-temperature case. The first order correction in $z_c^d$ is negative while the second order correction, though positive, is of negligible magnitude. Consequently, the overall impact of deformation on HEN at zero temperature is negative leading to a reduction in HEN. The behavior of deformation corrections in field theory (or equivalently cutoff corrections in the bulk) following the area law\footnote{Note that entanglement negativity also exhibits area law scaling as demonstrated in various systems such as finite-temperature quantum spin models and two-dimensional harmonic lattices \cite{DeNobili:2016nmj,PhysRevE.93.022128}.} is similar to the behavior of finite temperature corrections. This observed response of HEN to deformation indicates that deforming the theory destroys the quantum correlations between the subsystems, which leads to a diminished entanglement between them. This behavior is similar to the effects of increasing temperature which also reduces the entanglement between the subsystems.

\subsubsection{Small deformation and high-temperature regime: $ \tilde{\lambda}\ll l_{1,2}$ and $ 1/l_{1,2}\ll T$}

In contrast to the earlier subsection, we now have high-temperature limit $Tl_{1,2}\gg 1$ corresponding to subsystems $l_{1,2}$ while having the small deformation limit $\tilde{\lambda}\ll l_{1,2}$ in the theory. The holographic entanglement negativity at this regime for the mixed state configuration of adjacent subsystems can be obtained using \cref{HTSD} for $S(A_{1,2,12})$ in \cref{HEN-adjacent-formula} as follows:

\begin{equation}\label{adj-small-def-high-temp}
\begin{aligned}
 \mathcal{E} &=\frac{3R^{d-1}}{8G_N^{(d+1)}}\left(\frac{L}{z_c}\right)^{d-2}\Biggl[\frac{1}{d-2}-\frac{1}{4}\left(\frac{z_c}{z_h}\right)^d-\frac{3}{8(d+2)}\left(\frac{z_c}{z_h}\right)^{2 d} \Biggr]\\
 +&\frac{3R^{d-1}}{8G_N^{(d+1)}}\left(\frac{L}{z_h}\right)^{d-2}\Biggl[\hat{\mathcal{S}}-\sqrt{\frac{d-1}{2d}}(\epsilon_{l_1}+\epsilon_{l_2}-\epsilon_{l_1+l_2})
  +\frac{1}{2d}\left(\frac{z_c}{z_h}\right)^d
 +\frac{1}{8d}\left(\frac{z_c}{z_h}\right)^{2d}\Biggr]+...,
\end{aligned}
\end{equation}
where $\epsilon_{l_i}$ is given by \cref{Epsilon-high-temp} for different subsystems $A_i$. We observe that the corrections due to temperature and cutoff appear only through the dimensionless parameter $z_c/z_h$ and are independent of the subsystem sizes $l_i$. Since the theory is at high temperature, we see that the thermal contributions that are proportional to the volume $V=lL^{d-2}$ from \cref{HTSD} are completely subtracted at the leading order and leave us with a result where all of these corrections follow the area law behavior. This is expected as entanglement negativity provides an upper bound on distillable entanglement and should not involve volume-dependent thermal contributions at high temperatures.
The above result reduces to the following expression in the limit $z_c \to 0$

\begin{equation}
\begin{aligned}
 \mathcal{E}_{z_c\to 0} =\frac{3R^{d-1}}{8G_N^{(d+1)}}&\left(\frac{L}{z_c}\right)^{d-2}\left(\frac{1}{d-2}\right)+\frac{3R^{d-1}}{8G_N^{(d+1)}}\left(\frac{L}{z_h}\right)^{d-2}\Biggl[\hat{\mathcal{S}}-\sqrt{\frac{d-1}{2d}}\epsilon_d \Bigg\{\exp(-\sqrt{d(d-1)/2}\,\,\frac{l_1}{z_h})\\
 &+\exp(-\sqrt{d(d-1)/2}\,\,\frac{l_2}{z_h})-\exp(-\sqrt{d(d-1)/2}\,\,\frac{(l_1+l_2)}{z_h})\Bigg\}\Biggr]+...,
\end{aligned}
\end{equation}
where $z_c$ is the UV cutoff. It matches with the corresponding HEN for adjacent subsystems in the high-temperature limit \cite{Jain:2017xsu}.

\subsubsection{Large deformation and low-temperature regime: $l_{1,2} \ll \tilde{\lambda}$ and $T \ll 1/l_{1,2}$}

For this regime, we utilize \cref{LTLD} for $S(A_1)$, $S(A_2)$ and $S(A_1 \cup A_2)$ in \cref{HEN-adjacent-formula} to obtain the holographic entanglement negativity for adjacent subsystems as
\begin{equation}\label{adj-large-def-low-temp}
\begin{aligned}
  \mathcal{E}=&\frac{3R^{d-1}}{16G_N^{(d+1)}}\left(\frac{L}{z_c}\right)^{d-2}\Biggl[ -\frac{(d-1)^2}{24}\left(1-\left(\frac{z_c}{z_h}\right)^d\right)~\tilde{\mathcal{G}}_{3}^{\mathrm{adj}}\\
  +&\frac{(d-1)^3}{1920}(d+7)
\left(1+ 2 \frac{d-7}{d+7}\left(\frac{z_c}{z_h}\right)^d-\frac{3d-7}{d+7}\left(\frac{z_c}{z_h}\right)^{2d}\right)~\tilde{\mathcal{G}}_5^{\mathrm{adj}} \Bigg]+...,
\end{aligned}
\end{equation}
where $\tilde{\mathcal{G}}_n^{\mathrm{adj}} = z_c^{-n} \left( l_1^n+l_2^n-(l_1+l_2)^n\right)$.
As we observed from \cref{LTLD}, the leading contribution comes from the volume term due to finite cutoff (deformation) and scales as $S_A\sim V (1/z_c)^{d-1}$ in large deformation limit. This form is similar to volume term arising at high temperature where the entanglement entropy scales as $S_A\sim VT^{d-1}$. So, we expect that deformation acts as the destroyer of quantum correlations and introduces other degrees of freedom (or classical correlations) in the same way that raising the temperature does.
In the above result, we see that the volume contribution due to finite cutoff gets completely subtracted off from \cref{LTLD} at leading order in large deformation limit similar to the case of high temperature regime of previous subsection. Hence, the final result becomes proportional to the area of the entangling surface as expected since entanglement negativity captures only quantum correlations which are area dependent terms here. In this sense, the deformation and temperature affects the HEN in a similar way.

\subsection{Disjoint subsystems}
We now move to computation of the holographic entanglement negativity for the mixed state configuration of two disjoint subsystems at a finite temperature with deformation. This configuration consists of two long rectangular strips $A_1$ and $A_2$ separated by an another strip $A_s$ which are described by
\begin{equation}
x\equiv x^1=[-l_j/2,l_j/2],\quad x^i=[-L/2,L/2];\quad i=2,3,\dots,(d-1),\quad j=1,2,s,
\end{equation}
where $L$ is taken to be very large $L\gg l_1,l_2,l_s$ to neglect the corner effects.
The HEN between two disjoint subsystems can be computed using the following expression \cite{KumarBasak:2020viv}
\begin{equation}\label{EN_disjoint-formula}
\begin{aligned}
    \mathcal{E}=&\frac{3}{16G_N^{(d+1)}}(\mathcal{A}_{1s}+\mathcal{A}_{s2}-\mathcal{A}_{12s}-\mathcal{A}_s),\\
    &=\frac{3}{4}(S(A_1\cup A_s)+S(A_s \cup A_2)-S(A_1\cup A_2 \cup A_s)-S(A_s)),
    \end{aligned}
\end{equation}
where $\mathcal{A}_{ij}$ and $\mathcal{A}_{ijk}$ corresponds to the areas of the bulk RT surfaces for the subsystems $A_i\cup A_j$ and $A_i\cup A_j\cup A_k,$ respectively with $i=1,2,s$ and the expression in the second line can be obtained using the RT prescription \cite{Ryu:2006bv}.
Note that the above prescription is valid only in the proximity regime that is described by the condition that the separation between the two disjoint subsystems along the partitioning direction is much smaller than the lengths of the subsystems along that direction i.e. $l_s \ll l_{1,2}$. 
Since we have five parameters, namely $l_1,l_2,l_s, \tilde{\lambda}, T$ in the theory which leads to different limits involving these parameters to get the results in analytic form. Broadly we have three different limits of deformation parameters which are small deformation regime ($\tilde{\lambda} \ll l_s \ll l_{1,2}$), intermediate deformation regime ($l_s \ll \tilde{\lambda} \ll l_{1,2}$) and large deformation regime ($l_s \ll l_{1,2} \ll \tilde{\lambda}$). The imposition of distinct temperature limits on $l_1,l_2$ and $l_s$ in each of these cases leads to a plurality of different HEN results which is discussed in detail in the following subsections.

\subsubsection{Small deformation regime: $\tilde{\lambda} \ll l_s \ll l_{1,2}$}
The subsystem sizes in small deformation regime are such that $\frac{\tilde{\lambda}}{l_{1,2,s}} \ll 1$. Since we also have an additional temperature parameter $T$ in the field theory, we have three cases corresponding to small deformation regime  i.e., small-, intermediate-, and large-temperature regimes as discussed below.

\subsubsection*{Small deformation and low-temperature regime: $\tilde{\lambda} \ll l_s \ll l_{1,2}$ and $T \ll  1/l_{1,2,s}$}

In this regime, we have the small-temperature limit such that $Tl_{1,2,s} \ll  1$ as well as small deformation limit $\tilde{\lambda} \ll l_{1,2,s}$ corresponding to all the subsystems $l_{1,2,s}$. Now using \cref{LTSD} for $S(A_1\cup A_s)$, $S(A_2\cup A_s)$, $S(A_1\cup A_2\cup A_s),$ and $S(A_s)$ in \cref{EN_disjoint-formula}, the holographic entanglement negativity for the finite temperature and cutoff mixed state of two disjoint subsystems can be obtained as
\begin{equation}\label{EN-disj-LDLT}
\begin{aligned}
\mathcal{E} = \frac{3R^{d-1}}{8G_N^{(d+1)}}~\left(\frac{L}{z_c}\right)^{d-2}\Bigg[&\left(a_1+ a_5 \left(\frac{z_c}{z_h}\right)^d+a_9 \left(\frac{z_c}{z_h}\right)^{2d}\right) \mathcal{F}_{d-2}^{\mathrm{dis}}
  + \left(a_2+a_6 \left(\frac{z_c}{z_h}\right)^{d}\right) \mathcal{F}_{2d-2}^{\mathrm{dis}}
  + a_3~ \mathcal{F}_{3d-2}^{\mathrm{dis}}\Bigg]\\
  +\frac{3R^{d-1}}{8G_N^{(d+1)}}~\left(\frac{L}{z_h}\right)^{d-2}\Bigg[&\left(a_4+a_8 \left(\frac{z_c}{z_h}\right)^d\right) \mathcal{G}_2^{\mathrm{dis}}
  + a_7 ~\mathcal{G}_{d+2}^{\mathrm{dis}}\Bigg]+...,
    \end{aligned}
    \end{equation}
where the functions $\mathcal{F}_n^{\mathrm{dis}}$ and $\mathcal{G}_n^{\mathrm{dis}}$ are given by

\begin{equation}
  \begin{aligned}
  \mathcal{F}_n^{\mathrm{dis}} &= z_c^n \left(\frac{1}{(l_1+l_s)^n}+ \frac{1}{(l_2+l_s)^n} - \frac{1}{(l_1+l_2+l_s)^n}-\frac{1}{l_s^n}\right), \\ \mathcal{G}_n^{\mathrm{dis}} &= z_h^{-n} \left((l_1+l_s)^n+(l_2+l_s)^n - (l_1+l_2+l_s)^n-l_s^n\right). 
  \end{aligned}
\end{equation}
We observe that the result in \cref{{EN-disj-LDLT}} is cutoff independent in the limit $z_c\to 0,$ unlike the case of adjacent subsystems, and this behavior is similar to corresponding lower dimension results \cite{Malvimat:2018txq}. Interestingly if we take zero temperature and deformation limit ($T\to 0,z_c\to 0$) in \cref{{EN-disj-LDLT}}, we recover the results for the zero temperature mixed state of two disjoint subsystems in field theory dual to pure AdS$_{d+1}$ \cite{KumarBasak:2020viv}. 
In addition to the contribution from AdS$_{d+1}$ vacuum, the remaining terms are the corrections due to both temperature and deformation parameter. All of these corrections are  proportional to the area of the subsystem $L^{d-2}$, thus following the area law. The sum of all these corrections due to finite temperature and cutoff is negative, so their combined effect decreases the HEN. 

On taking the limit $z_c \to 0$ in \cref{EN-disj-LDLT}, the HEN for two disjoint subsystems reduces to the following expression

\begin{equation}
\begin{aligned}
\mathcal{E}_{z_c \to 0} = \frac{3R^{d-1}}{8G_N^{(d+1)}}~L^{d-2}\Bigg[& a_1 \left(\frac{1}{(l_1+l_s)^{d-2}}+ \frac{1}{(l_2+l_s)^{d-2}} - \frac{1}{(l_1+l_2+l_s)^{d-2}}-\frac{1}{l_s^{d-2}}\right)\\
  &+\left(\frac{1}{z_h}\right)^{d-2}\left(a_4 \mathcal{G}_2^{\mathrm{dis}}
  + a_7 ~\mathcal{G}_{d+2}^{\mathrm{dis}}\right)\Bigg].
    \end{aligned}
    \end{equation}
The above result matches with corresponding entanglement negativity for disjoint subsystems in the small temperature limit up to order $1/z_h^d$ \cite{KumarBasak:2020viv}. This matching serves as a consistency check for our computations.

\subsubsection*{Small deformation and intermediate-temperature regime: $\tilde{\lambda} \ll  l_s \ll l_{1,2}$ and $1/l_{1,2} \ll T \ll 1/l_s$}

In this regime, we have small deformation and high-temperature limit corresponding to subsystems $l_{1,2}$ but low temperature with small deformation limit for subsystem $l_s$. So, we use \cref{HTSD} for $S(A_1\cup A_s)$, $S(A_2\cup A_s)$, $S(A_1\cup A_2\cup A_s)$ and \cref{LTSD} for $S(A_s)$ in \cref{EN_disjoint-formula} to obtain the HEN as

\begin{equation}
\begin{aligned}
\mathcal{E} =\frac{3R^{d-1}}{8G_N^{(d+1)}}\left(\frac{L}{z_h}\right)^{d-2}&\Biggl[\frac{l_s}{2z_h}+\hat{\mathcal{S}}-\sqrt{\frac{d-1}{2d}}\left(\epsilon_{l_1+l_s}+\epsilon_{l_2+l_s} - \epsilon_{l_1+l_2+l_s}\right)+\frac{1}{2d}\left(\frac{z_c}{z_h}\right)^d+\frac{1}{8d}\left(\frac{z_c}{z_h}\right)^{2d}\\
&-\left(a_4+a_8 \left(\frac{z_c}{z_h}\right)^d\right) \left(\frac{l_s}{z_h}\right)^2- a_7 \left(\frac{l_s}{z_h}\right)^{d+2} \Biggr]\\
-\frac{3R^{d-1}}{8G_N^{(d+1)}}\left(\frac{L}{l_s}\right)^{d-2}&\Biggl[\left(a_1+ a_5 \left(\frac{z_c}{z_h}\right)^d+a_9 \left(\frac{z_c}{z_h}\right)^{2d}\right)+\left(a_2+a_6 \left(\frac{z_c}{z_h}\right)^{d}\right) \left(\frac{z_c}{l_s}\right)^d
+a_3 \left(\frac{z_c}{l_s}\right)^{2d} \Biggr]+...,
\end{aligned}
\end{equation}
where $\epsilon_{l_i}$ given by \cref{Epsilon-high-temp} corresponds to different subsystem sizes $l_i$. Here, the corrections to HEN due to deformation and temperature follows the area law behavior as can be seen from terms proportional to the dimensionless parameter $(\frac{z_c}{z_h})$. The first term in the above result is the volume thermal term and its presence can be attributed to fact that we are in small-temperature regime corresponding to subsystem $A_s$, hence it does not get completely subtracted off. However, the thermal contributions from subsystems $A_1$ and $A_2$ get completely subtracted since they are in large temperature limit $Tl_{1,2}\gg 1$. The rest of the terms including $\hat{\mathcal{S}}$ give negative contribution, hence decreasing the holographic entanglement negativity.

\subsubsection{Intermediate deformation regime: \textbf{$l_s \ll \tilde{\lambda} \ll l_{1,2}$}}

This regime is characterized by two distinct temperature limits: the small-temperature regime $(T \ll 1/l_{1,2,s})$ and an intermediate-temperature regime $(1/l_{1,2} \ll T \ll 1/l_s)$ which are discussed in the following subsections.

\subsubsection*{Intermediate deformation and low-temperature regime: $l_s \ll \tilde{\lambda} \ll l_{1,2}$ and $T \ll 1/l_{1,2,s}$}

In this regime, we have the low-temperature limit corresponding to all subsystem $l_{1,2,s}$ such that $T l_{1,2,s} \ll 1$, for intermediate deformation of the theory $l_s \ll \tilde{\lambda} \ll l_{1,2}$. So, the holographic entanglement negativity for disjoint subsystems in this regime may be obtained using \cref{LTSD} for $S(A_1\cup A_s)$,and $S(A_2\cup A_s)$, $S(A_1\cup A_2\cup A_s)$ and \cref{LTLD} for $S(A_s)$ in \cref{EN_disjoint-formula} as

\begin{equation}\label{EN-disj-intm-def-low-temp}
  \begin{aligned}
\mathcal{E} =&\frac{3R^{d-1}}{8G_N^{(d+1)}}\left(\frac{L}{z_c}\right)^{d-2}\Biggl[\frac{1}{d-2}-\frac{1}{4}\left(\frac{z_c}{z_h}\right)^{d}-\frac{3}{8(d+2)} \left(\frac{z_c}{z_h}\right)^{2d}-\left(\frac{l_s}{2z_{c}}\right)\\&+\frac{(d-1)^2}{6}\left(1-\left(\frac{z_c}{z_h}\right)^{d}\right)\left(\frac{l_s}{2z_{c}}\right)^{3}\\
 &-\frac{(d-1)^3}{120}(d+7)
\left(1+ 2 \frac{d-7}{d+7}\left(\frac{z_c}{z_h}\right)^d-\frac{3d-7}{d+7}\left(\frac{z_c}{z_h}\right)^{2d}\right)~\left(\frac{l_s}{2 z_c}\right)^5\\
  &+\left(a_1 + a_5 \left(\frac{z_c}{z_h}\right)^d + a_9 \left(\frac{z_c}{z_h}\right)^{2d}\right)\hat{\mathcal{F}}_{d-2}^{\mathrm{dis}}
  +\left(a_2 + a_6 \left(\frac{z_c}{z_h}\right)^d\right)\hat{\mathcal{F}}_{2d-2}^{\mathrm{dis}}+ a_3~ \hat{\mathcal{F}}_{3d-2}^{\mathrm{dis}}\Biggr]\\
  & +\frac{3R^{d-1}}{8G_N^{(d+1)}}\left(\frac{L}{z_h}\right)^{d-2}\Biggl[\left(a_4 + a_8 \left(\frac{z_c}{z_h}\right)^d\right) \hat{\mathcal{G}}_{2}^{\mathrm{dis}} + a_7 \hat{\mathcal{G}}_{d+2}^{\mathrm{dis}}\Biggr]+...,
\end{aligned}
\end{equation}
where the functions $\hat{\mathcal{F}}_{n}^{\mathrm{dis}}$ and $\hat{\mathcal{G}}_{n}^{\mathrm{dis}}$ are given by

\begin{equation}
    \begin{aligned}
    \hat{\mathcal{F}}_{n}^{\mathrm{dis}}&= z_c^n \left(\frac{1}{(l_1+l_s)^n}+\frac{1}{(l_2+l_s)^n}- \frac{1}{(l_1+l_2+l_s)^n}\right),\\
    \hat{\mathcal{G}}_{n}^{\mathrm{dis}}&=z_h^{-n} \left((l_1+l_s)^n+(l_2+l_s)^n - (l_1+l_2+l_s)^n\right).
    \end{aligned}
\end{equation}
We see that both temperature and deformation parameters contribute to corrections in the HEN separately similar to the small deformation regime. Notably, both types of corrections lead to a decrease in  HEN indicating a reduction in the HEN between the subsystems. Note that the above result reduces to HEN for adjacent subsystems in the adjacent limit $l_s\to 0$ as discussed in \cref{EN-ajd-small-def-low-temp}.

By taking the limit as the temperature $T\to 0$, we can effectively eliminate thermal fluctuations and focus solely on the impact of the deformation parameter on the HEN at a finite cutoff value. The holographic entanglement negativity from \cref{EN-disj-intm-def-low-temp} at $T\to 0$ reduces to

\begin{equation}
  \begin{aligned}
\mathcal{E}_{T\to 0} =\frac{3R^{d-1}}{8G_N^{(d+1)}}\left(\frac{L}{z_c}\right)^{d-2}&\Biggl[\frac{1}{d-2}-\frac{l_s}{2z_{c}}+\frac{(d-1)^2}{6}\left(\frac{l_s}{2z_{c}}\right)^{3}-\frac{(d-1)^3}{120}(d+7)\left(\frac{l_s}{2 z_c}\right)^5\\
&+a_1\hat{\mathcal{F}}_{d-2}^{\mathrm{dis}}
  +a_2\hat{\mathcal{F}}_{2d-2}^{\mathrm{dis}}+ a_3 \hat{\mathcal{F}}_{3d-2}^{\mathrm{dis}}\Biggr]+...
\end{aligned}
\end{equation}
Since we are in the high deformation limit $l_s \ll \tilde{\lambda}$ for  subsystem $A_s$ in the intermediate deformation regime, we have a presence of volume term arising from finite cutoff (or deformation) and it does not cancel as subsystems $A_{1,2}$ are in the small deformation regime. However its contribution is negative and the HEN decreases on increasing the deformation parameter. So, deforming the system causes the decrease in quantum information shared between the disjoint subsystems. This behavior resembles the impact of temperature on entanglement negativity where an increase in temperature similarly leads to a reduction in its value.

\subsubsection*{Intermediate deformation and intermediate-temperature regime: $l_s \ll \tilde{\lambda} \ll l_{1,2}$ and $1/l_{1,2} \ll T \ll 1/l_s$}
In contrast to the earlier subsection, we now have small deformation and high-temperature limit corresponding to subsystems $l_{1,2},$ while having a large deformation and small-temperature limit for the subsystem $l_s$. So, we now use \cref{HTSD} for $S(A_1\cup A_s)$, $S(A_2\cup A_s)$, $S(A_1\cup A_2\cup A_s)$ and \cref{LTLD} for $S(A_s)$ in \cref{EN_disjoint-formula} to obtain the HEN for mixed state configuration of two disjoint subsysytems in this regime as
\begin{equation}
\begin{aligned}
 \mathcal{E}=\frac{3R^{d-1}}{8G_N^{(d+1)}}\left(\frac{L}{z_c}\right)^{d-2}&\Biggl[ \frac{1}{d-2}+\left(\left(\frac{z_c}{z_h}\right)^{d-1}-1\right)\left(\frac{l_s}{2z_c}\right)
  +\frac{(d-1)^2}{6}\left(1-\left(\frac{z_c}{z_h}\right)^d\right)\left(\frac{l_s}{2z_c}\right)^3 \\
  &-\frac{(d-1)^3}{120}(d+7)
\left(1+ 2 \frac{d-7}{d+7}\left(\frac{z_c}{z_h}\right)^d-\frac{3d-7}{d+7}\left(\frac{z_c}{z_h}\right)^{2d}\right)~\left(\frac{l_s}{2 z_c}\right)^5\\
  &-\frac{1}{4}\left(\frac{z_c}{z_h}\right)^d-\frac{3}{8(d+2)}\left(\frac{z_c}{z_h}\right)^{2d}
  \Biggr]\\ 
  +\frac{3R^{d-1}}{8G_N^{(d+1)}}\left(\frac{L}{z_h}\right)^{d-2}&\Biggl[\hat{\mathcal{S}}-\sqrt{\frac{d-1}{2d}}\left(\epsilon_{l_1+l_s}+\epsilon_{l_2+l_s} - \epsilon_{l_1+l_2+l_s} \right)+\frac{1}{2d}\left(\frac{z_c}{z_h}\right)^{d}+\frac{1}{8d}\left(\frac{z_c}{z_h}\right)^{2d}\Biggr]+...
  \end{aligned}
\end{equation}
The first line in the above equation has a volume term arising from deformation similar to the previous subsection and has overall negative contribution. In the final line, the thermal volume terms associated with subsystems $A_{1,2}$ are eliminated owing to the high-temperature limit imposed on these subsystems. In the adjacent limit $l_s\to 0$, we again observe that the above result simplifies to the HEN for adjacent subsystems at small deformation and high-temperature regime as given in \cref{adj-small-def-high-temp}.

\subsubsection{Large deformation regime: $l_s \ll l_{1,2} \ll \tilde{\lambda}$}

In this regime, we only have low-temperature limit corresponding to each subsystems i.e. $T \,l_{1,2,s}\ll 1$ as described in the dual picture as follows. In this particular case, the turning point $z_*$ which is associated with each subsystem is very close to the cutoff surface $z_c$. As $z_c$ is itself situated far from the horizon $z_h$, this regime restricts the analysis to the low-temperature limit. The holographic entanglement negativity for two disjoint subsystems in this regime may now be obtained using \cref{LTLD} for $S(A_{1,2,s})$ in \cref{EN_disjoint-formula} as

\begin{equation}
\begin{aligned}
  \mathcal{E}=\frac{3R^{d-1}}{16G_N^{(d+1)}}\left(\frac{L}{z_c}\right)^{d-2}\Biggl[& -\frac{(d-1)^2}{24}\left(1-\left(\frac{z_c}{z_h}\right)^d\right)~\tilde{\mathcal{G}}_{3}^{\textrm{dis}}
  +\frac{(d-1)^3}{1920}(d+7)
\Bigg(1+ 2 \frac{d-7}{d+7}\left(\frac{z_c}{z_h}\right)^d\\
&-\frac{3d-7}{d+7}\left(\frac{z_c}{z_h}\right)^{2d}\Bigg)~\tilde{\mathcal{G}}_5^{\textrm{dis}} \Biggr]+...,
\end{aligned}
\end{equation}
where $\tilde{\mathcal{G}}_n^{\textrm{dis}} = z_c^{-n} \left( (l_1+l_s)^n+(l_2+l_s)^n-(l_1+l_2+l_s)^n-l_s^n\right)$ and ellipses denotes the higher order terms. We observe that in the large deformation limit, the volume-dependent terms arising from the finite cutoff in \cref{LTLD} subtracts between the two disjoint subsystems which leads to the HEN proportional to the area ($L^{d-2}$) of the entangling surface. This behavior is similar to the cancellation of thermal volume terms in the high-temperature regime as observed for adjacent subsystems. Thus, we conclude that deformation and temperature exhibit analogous effects on HEN. If we take the adjacent limit $l_s\to 0$, the above result reduces to the corresponding HEN for the adjacent subsystems described in \cref{adj-large-def-low-temp}. It is important to note that the above result differs from the behavior of mutual information between disjoint subsystems in the large deformation regime \cite{Ebrahim:2023ush}. In the case of mutual information, a volume dependent term  due to finite cutoff is present as expected, since mutual information measures total correlation including classical correlations. In contrast entanglement negativity as a measure of quantum entanglement does not exhibit such volume-dependent contributions at large deformation limit.

\section{Summary and Discussion}\label{summary}

To summarize, we have studied mixed state entanglement measures such as EoP, which is dual to the EWCS and entanglement negativity in a higher-dimensional deformed field theory at finite temperature. The deformation of field theory by a generalized $T\bar{T}$ operator is equivalent to introducing a nonzero cutoff in the bulk. The bulk dual to $T\bar{T}$ deformed CFT$_d$ is described by a cutoff AdS$_{d+1}$ geometry.
So, the consequence of high temperature or large deformation is a deeper probing into the bulk. 
We analyzed the entanglement structure of a mixed state by computing the EWCS and entanglement negativity holographically at different limits of temperature and deformation parameter in the deformed geometry. We note that the impact of deformation and temperature on the mixed state entanglement measures we investigated is similar to the behavior observed for entanglement entropy and mutual information in \cite{Ebrahim:2023ush}. Importantly, deformation decreases the entanglement between the subsystems.

We first computed the EWCS in different regimes of temperature and deformation parameter in the present setup. Our study of the EWCS reveals that it exhibits area dependence in the limiting cases in contrast to the entanglement entropy which scales with the volume at high temperatures and deformations. Furthermore, we find that deformation significantly influences the entanglement structure of the boundary field theory at both finite and zero temperatures. Notably, our results showed a reduction in correlation with increasing deformation as indicated by the decreasing behavior of the EWCS with respect to the deformation.  In low and intermediate deformation we observe that the EWCS follows the area law at both high and low thermal limits. However at large deformation the area terms are absent as can be seen from \cref{hdf}. 
It is important to note that in the low deformation EWCS diverges as the distance $D \to 0$, however it remains finite for intermediate and the large deformation. This behavior is similar to mutual information observed in \cite{Ebrahim:2023ush}. Our results aligned with previously reported results in the appropriate parameter limits validating the consistency of our analysis.

Subsequently we investigated the effect of deformation on holographic entanglement negativity for mixed state configuration of two adjacent and disjoint subsystems. We observed that corrections to leading order terms obey area law in all regimes.
In the high deformation limit, we observed a cancellation between the volume terms coming due to finite cutoff (or deformation) leaving the HEN to be dependent on the area of the entangling surface. This behavior closely resembles that observed at high temperature where a similar cancellation of volume thermal terms occurs \cite{Jain:2017xsu,KumarBasak:2020viv}. Furthermore, all temperature and deformation correction terms contribute negatively, effectively decreasing HEN and consequently reducing the entanglement between the subsystems. We observe that the effect of raising temperature or the deformation parameter appears to bring other degrees of freedom in the correlation between subsystems. This behavior leads us to conclude that deformation and temperature have similar effects on HEN as well as on the EWCS. Interestingly our results exactly match those reported in the literature in the limit of vanishing deformation parameter which provides a consistency check for our results. We see that increasing the deformation or temperature leads to deeper probing into the bulk, which introduces more fluctuations. As a result, the entanglement between the subsystems decreases.

Since the EWCS has been demonstrated to be dual to mixed state entanglement measures like entanglement of purification, reflected entropy, entanglement negativity and odd entropy, our analysis of it can provide novel insights into the intricate structure of mixed state entanglement in deformed field theories at finite temperature. 
In our study, we only considered the striplike subsystem, however investigating mixed state measures for a generic subsystem of geometries in higher dimensions would be interesting. It would also be interesting to explore various mixed state measures in higher-dimensional nonrelativistic framework, particularly in the presence of deformation. At various occasions it is observed that the different kind of deformation, anisotropy greatly affect the chaotic nature of the theory \cite{Karan:2023hfk, Chakrabortty:2022kvq,Jahnke:2017iwi, Sil:2020jhr}. It would be interesting to explore the chaotic behavior in the presence of finite cutoff for generic dimensions. We leave these open problems for future work.

\section*{ACKNOWLEDGMENTS}

We would like to thank Debanjan Karan for collaboration on this project in its early stages. H.P acknowledges the support of this work by NCTS.

\begin{appendices}

	\section{Numerical coefficients}\label{Num-coeff}
		The coefficients $a_i$ used in \cref{LTSD} are given by \cite{Ebrahim:2023ush}

\begin{equation}
\begin{aligned}
\label{as}
  &a_1 \equiv \frac{(2 c_0)^{d-1}}{2(2-d)},\\
  &a_2 \equiv \frac{1}{2d}(2 c_0)^{2(d-1)},\\
  &a_3 \equiv \frac{\left(1-d\right)}{d^2}(2c_0)^{3(d-1)},\\
  &a_4 \equiv \frac{\left(d-1\right)c_1}{4(2 c_0)^2},\\
  &a_5 \equiv \frac{\left(d-1\right)c_1(2 c_0)^{d-3}}{d},\\
  &a_6 \equiv \frac{(2c_0)^{2(d-1)}}{8d}-\frac{\left(4d^2-7d+3\right)c_1 (2{c_0})^{2(d-2)}}{d^2},\\
  &a_7 \equiv \frac{3 \left(d-1\right) c_2}{8\left(d+2\right)(2c_0)^{d+2}}-\frac{\left(d-1\right) {c_1}^2}{4(2 c_0)^{d+3}},\\
  &a_8 \equiv \frac{\left(2d^2-d+3\right) {c_1}^2}{2d(2c_0)^4}-\frac{3c_2}{4d(2c_0)^3},\\
  &a_9 \equiv \frac{(2 c_0)^{d-5}}{4 d^2}\bigg(4d\left(2d-1\right){c_0}^2 c_1 - 2\left(9d^3-36d^2+19d-12\right){c_1}^2-6\left(5d^2-2d+3\right)c_0 c_2 \bigg).
    \end{aligned}
    \end{equation}

\end{appendices}

\bibliographystyle{JHEP}
\bibliography{TTbar}

\end{document}